\documentclass[a4paper,12pt]{article}        
\usepackage{geometry}           
\usepackage[T2A]{fontenc}
\usepackage[utf8]{inputenc}
\usepackage[english]{babel} 
\usepackage{amstext,amsmath,amssymb}            
\usepackage{bm}                 
\usepackage[pdftex]{graphicx}   
                                
\usepackage{float}
\usepackage{amsfonts}           
\usepackage{indentfirst}        
\usepackage{cite}               
\usepackage{multirow}           
\usepackage{array}              
\graphicspath{{figures/}}

\begin{document}

\begin{titlepage}
\author{Butler A.P.$^{a,b}$, Butler P.H.$^{b,c}$, Bell S.T.$^c$, Chelkov G.A.$^{d,e}$, Dedovich D.V.$^d$, \\ 
   Demichev M.A.$^d$, Elkin V.G.$^d$, Gostkin M.I.$^d$, Kotov S.A.$^d$, Kozhevnikov D.A.$^d$, \\
   Kruchonak U.G.$^d$, Nozdrin A.A.$^d$, Porokhovoy S.Yu.$^d$, Potrap I.N.$^d$, \\
   Smolyanskiy P.I.$^{d,}$\thanks{E-mail: smolyanskiy@jinr.ru}, Zakhvatkin M.M.$^d$, Zhemchugov A.S.$^d$ \\
  {\small {\it \llap{$^a$}University of Otago, Christchurch, NZ }} \\ 
  {\small {\it \llap{$^b$}University of Canterbury, Christchurch, NZ }} \\
  {\small {\it \llap{$^c$}MARS Bioengineering, 29a Clyde Road, Christchurch, NZ}} \\ 
  {\small {\it \llap{$^d$}Joint institute for nuclear research, Dubna, Russia } } \\
  {\small {\it \llap{$^e$}Tomsk state university, }} \\
  {\small {\it \llap{} Educational-research-innovation center ``Semiconductor sensors'', Tomsk, Russia  }}}
   
\title{Measurement of the energy resolution and calibration of hybrid pixel detectors with GaAs:Cr sensor and Timepix readout chip}

\end{titlepage}

\maketitle

\setcounter{page}{1}            




\section{Annotation}
This paper describes an iterative method of per-pixel energy calibration of hybrid pixel detectors with GaAs:Cr sensor and Timepix readout chip. A convolution of precisely measured spectra of characteristic X-rays of different metals with the resolution and the efficiency of the pixel detector is used for the calibration. The energy resolution of the detector is also measured during the calibration. The use of per-pixel calibration allows to achieve a good energy resolution of the Timepix detector with GaAs:Cr sensor: 8\% and 13\% at 60 keV and 20 keV, respectively.
\section{Introduction}

In recent years hybrid pixel semiconductor detectors have been finding more use in various fields of science where high spatial resolution and low noise are required from radiation detectors. In high-energy physics such detectors are used for registering particle tracks, in geology -- for X-ray radiography in studies of the internal structure of different samples, in biology and medicine -- for computer tomography. In most of these hybrid semiconductor detectors sensing elements are made of silicon (Si, Z=14), which has both undeniable advantages and considerable disadvantages in comparison with other semiconductors. In particular, Si has a very low efficiency of detection of photons with energies greater than 20 keV\footnote{This parameter is especially important for computer tomography of dense tissue and minerals.} and insufficient resistance to radiation. So, heavier alternatives to silicon such as gallium arsenide (GaAs, Z=31,33) and cadmium telluride (CdTe, Z=48,52) have increasingly being used as sensor material for hybrid pixel detectors.

The JINR Laboratory of Nuclear Problems together with the Tomsk State University, Medipix collaboration~\cite{medipix}, Czech Technical University (Prague) and the research center DESY (Hamburg), is working on developing and testing hybrid pixel detectors based on the Timepix readout chip~\cite{timepix} with sensor matrix of chromium-doped gallium arsenide (GaAs:Cr)~\cite{aiz}. This publication describes the latest results of these studies.

\section{Measurement of X-ray spectra with Canberra detector}

\begin{figure}[ht]
\center{\includegraphics[width=0.8\linewidth]{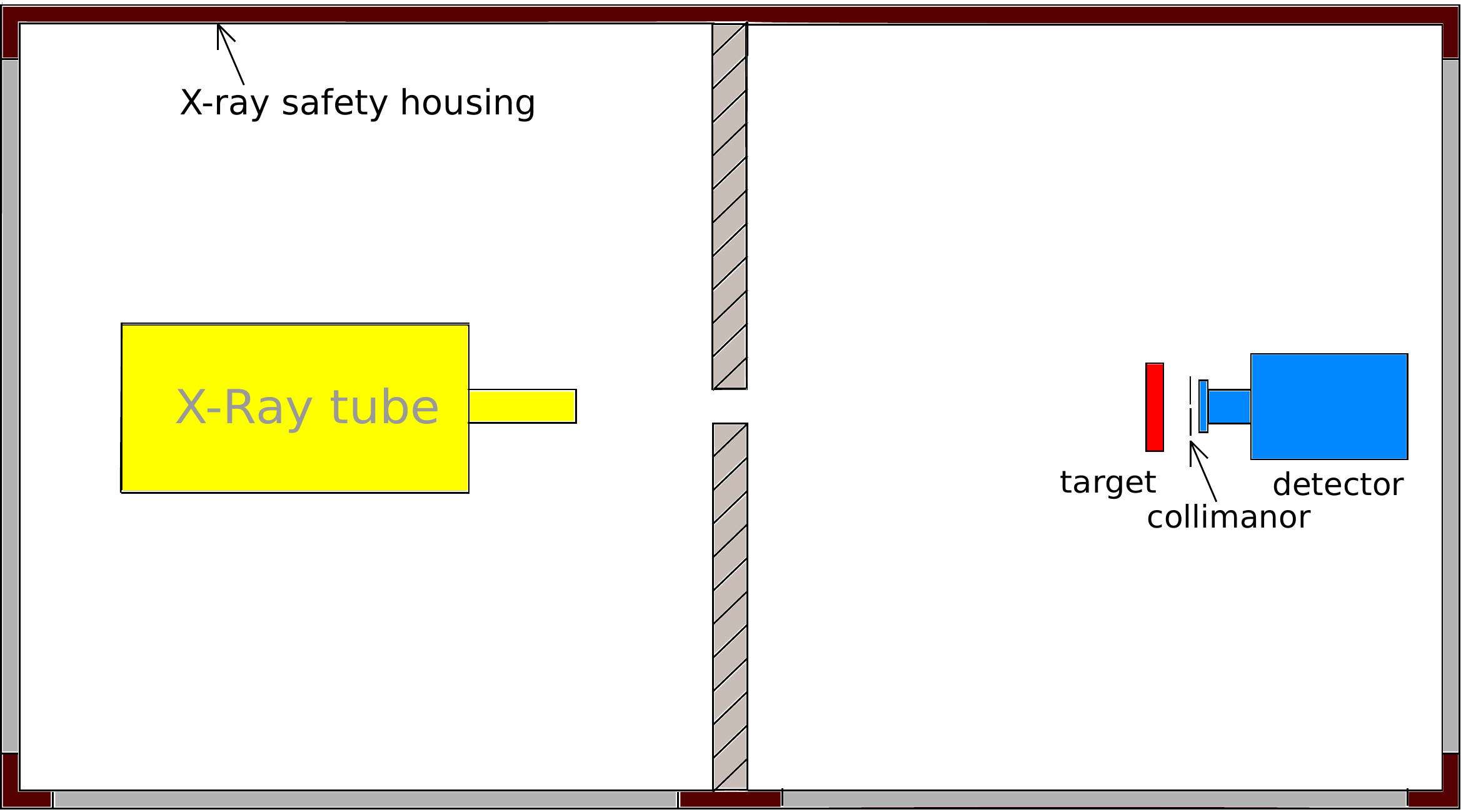}}
\caption{\footnotesize Layout of the experimental setup for measuring of X-ray spectra.
\label{Setup}}
\end{figure}

Schematic diagram of the experimental setup for performing spectral measurements is shown in Fig.~\ref{Setup}. The lead collimator placed in front of the detector window to prevent overloading in measurements of direct spectra had 5 mm thickness and a hole of 0.8 mm diameter in the middle. In the measurements we used a set of targets made of different metals from nickel to lead.

As X-ray sources we used microfocus X-ray tubes RAP-150MN~\cite{rap150} and SB120~\cite{sb120}. Their main characteristics are presented in Table~\ref{XrayTubesPars}. It is important to note, that in RAP-150MN tube emitted X-ray photons pass through the anode (made of aluminum coated with tungsten), while in SB120 tube X-ray photons are emitted directly from the anode (made of copper coated with tungsten) without passing through any material except the glass housing.

\begin{table}[h]
\caption{\label{XrayTubesPars}Basic properties of РАП-150МН and SB120 X-ray tubes}
\begin{center}
  \begin{tabular}{|l|c|c|} \hline
                      & RAP-150MN & SB120 \\ \hline
    Anode voltage, кV & $20 \div 150$ & $60 \div 120$ \\ \hline  
    Anode current, mkА & $20 \div 100$ & $10 \div 350$ \\ \hline
    Focal spot size, mkm & 60 & 70  \\ \hline
\end{tabular}
\end{center}
\end{table}

Energy spectra were measured with a precision $\gamma$-spectrometer Canberra GL0515R~\cite{canberra}. This detector has a high energy resolution in the range of X-ray radiation (15-300 keV) and is factory calibrated with high precision for the energy scale. Main characteristics of the Canberra HPGe detector are presented in Table ~\ref{CanberraPars}.

\begin{table}[h]
\caption{\label{CanberraPars}Technical data on Canberra GL0515R detector}
\begin{center}
  \begin{tabular}{|c|c|} \hline
    Active area, mm$^2$& 500 \\ \hline  
    Thickness, mm & 15 \\ \hline
    Thickness of Be window, mm & 0.05  \\ \hline
    Resolution ($\sigma$) at 5.9 keV, keV & 0.1 \\ \hline
    Resolution ($\sigma$) at 122 keV, keV & 0.23 \\ \hline
\end{tabular}
\end{center}
\end{table}


\subsection{Emission spectra of RAP-150MN and SB120 X-ray tubes}

\begin{figure}[p]
\center{\includegraphics[width=\linewidth]{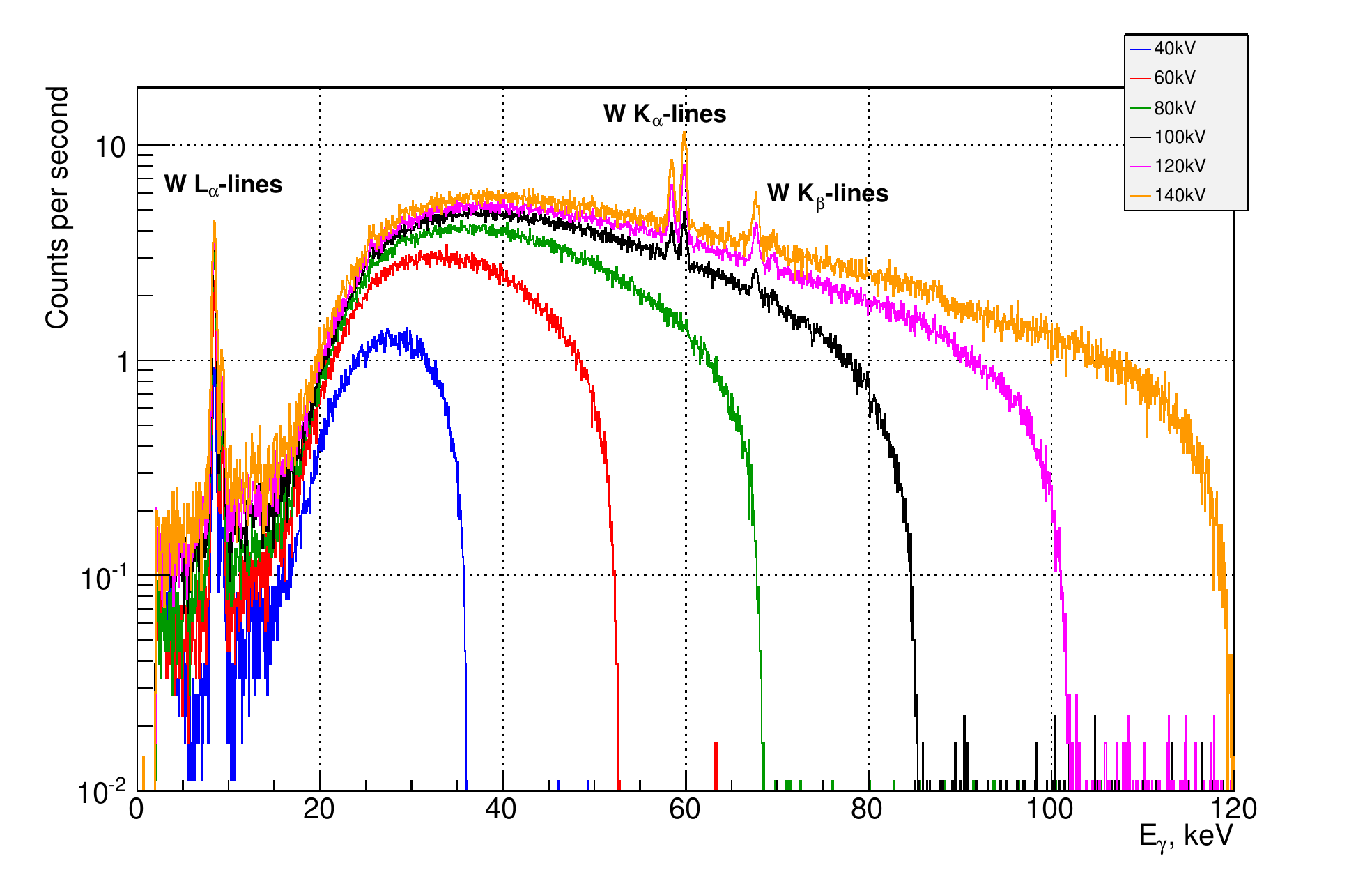}}
\caption{\footnotesize Emission spectra of RAP-150MN X-ray tube for different anode voltage. \label{RAP150Spectrum_log}}
\end{figure}

\begin{figure}[p]
\center{\includegraphics[width=\linewidth]{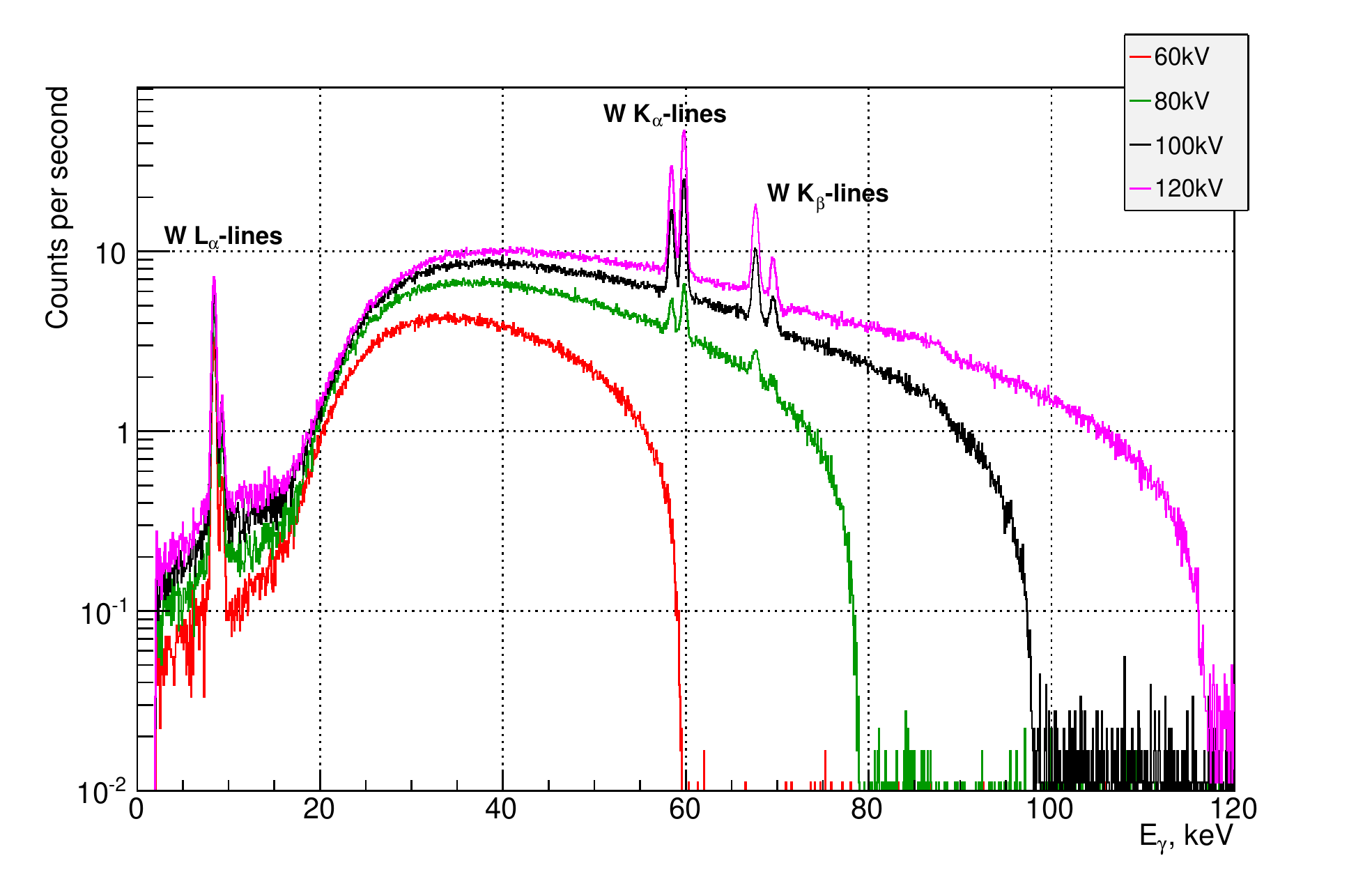}}
\caption{\footnotesize Emission spectra of SB120 X-ray tube for different anode voltage. \label{SB120Spectrum_log}}
\end{figure}

To get an idea of the energy distribution of the radiation intensity of the X-ray tubes RAP-150MN and SB120, their energy spectra were measured in the operating range of their anode voltage (for SB120 it is much narrower than for RAP-150MN). The emission spectra, normalized to the exposure time, at different anode voltages for the X-ray tubes RAP-150MN and SB120 are shown, respectively, in Fig.~\ref{RAP150Spectrum_log} and Fig.~\ref{SB120Spectrum_log}. As can be seen from the figures, the edge of the X-ray spectrum for the RAP-150MN tube does not correspond to the anode voltage set by the control unit, while for the SB120 tube the edge of the spectrum corresponds strictly to the set anode voltage.\footnote{It seems that the real accelerating voltage on the anode of the RAP-150MN tube differs from the voltage set by the control unit.} Since the X-ray tubes RAP-150MN and SB120 have anodes coated with tungsten, peaks corresponding to $K_{\alpha}$-,$K_{\beta}$- and $L_{\alpha}$-lines of tungsten characteristic emission are clearly distinguished in their continuous spectra. Events in the spectra ``tails'', with energies higher then the set anode voltage, are explained by superposition of signals from several photons.


\subsection{Spectra of characteristic X-ray emission from various metals}

\begin{figure}[htb]
\center{\includegraphics[width=0.8\linewidth]{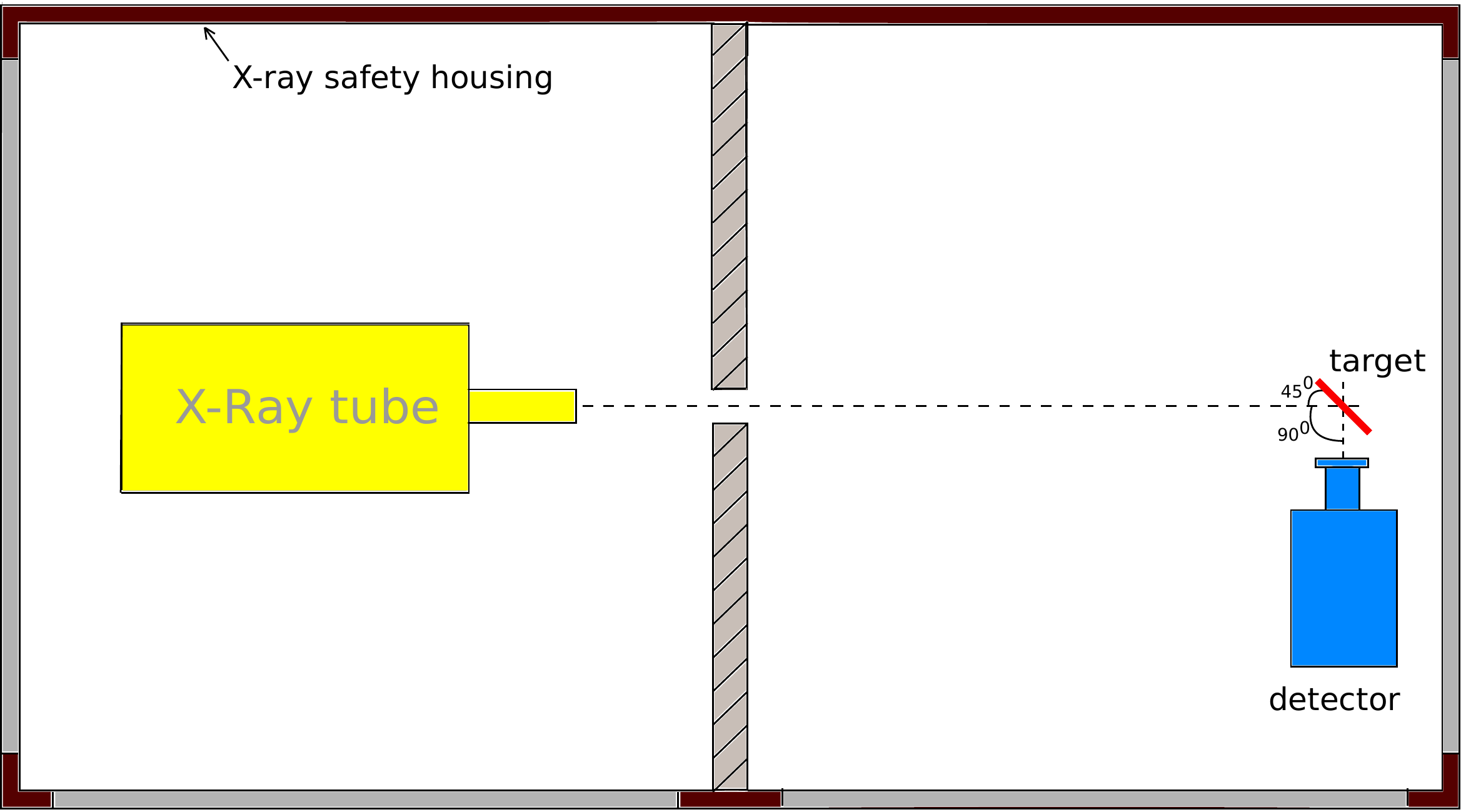}}
\caption{\footnotesize Layout of the experimental setup for measuring of X-ray spectra in ``reflection'' geometry.
 \label{Setup_2}}
\end{figure}

For measurements of characteristic emission spectra of the metals listed in Table~\ref{MetalsKLines} with the HPGe detector Canberra, the RAP-150MN X-ray tube was used. The measurements were performed in two geometries: ``transmission'' geometry when the target (foils of different metals) is placed between the emitter and the detector (Fig.\ref{Setup}), and ``reflection'' geometry when the target is placed at the angle of $45^{\circ}$ to the emitter axis and the detector is place at the angle of $90^{\circ}$ and off the emitter axis (Fig.\ref{Setup_2}). It should be noted that there is no collimator in front of the detector in the second case.

\begin{table}[htb]
\caption{\label{MetalsKLines}Energy of the characteristic emission of different metals}
\begin{center}
  \begin{tabular}{|c|c|c|c|c|c|c|c|} \hline
    Element & $K_{\alpha_2}$, keV & $K_{\alpha_1}$, keV & $K_{\beta_1}$, keV & $K_{\beta_2}$, keV & $K_{\beta_3}$, keV & $K_{\beta_4}$, keV & $K_{\beta_5}$, keV \\ \hline  
    $^{28}$Ni & 7.46 & 7.48 & 8.27 & -- & 8.27 & -- & 8.33 \\
    $^{30}$Zn & 8.62 & 8.64 & 9.57 & -- & 9.57 & -- & 9.65 \\ 
    $^{40}$Zr & 15.69 & 15.78 & 17.68 & 17.97 & 17.65 & 17.97 & 17.82 \\
    $^{42}$Mo & 17.38 & 17.48 & 19.61 & 19.97 & 19.59 & 19.97 & 19.77 \\
    $^{45}$Rh & 20.07 & 20.22 & 22.72 & 23.17 & 22.70 & 23.17 & 22.91 \\
    $^{48}$Cd & 22.98 & 23.17 & 26.10 & 26.64 & 26.06 & 26.64 & 26.30 \\
    $^{49}$In & 22.98 & 23.17 & 26.10 & 26.64 & 26.06 & 26.64 & 26.30 \\
    $^{50}$Sn & 25.04 & 25.27 & 28.49 & 29.11 & 28.44 & 29.11 & 28.71 \\
    $^{73}$Ta & 56.28 & 57.54 & 65.22 & 67.01 & 64.95 & 66.95 & 65.24 \\
    $^{74}$W & 57.98 & 59.32 & 67.24 & 69.10 & 66.95 & 69.03 & 67.65 \\
    $^{82}$Pb & 72.81 & 74.97 & 84.94 & 87.36 & 84.45 & 87.24 & 85.42 \\ \hline
\end{tabular}
\end{center}
\end{table}

\begin{figure}[htb]
\center{\includegraphics[width=\linewidth]{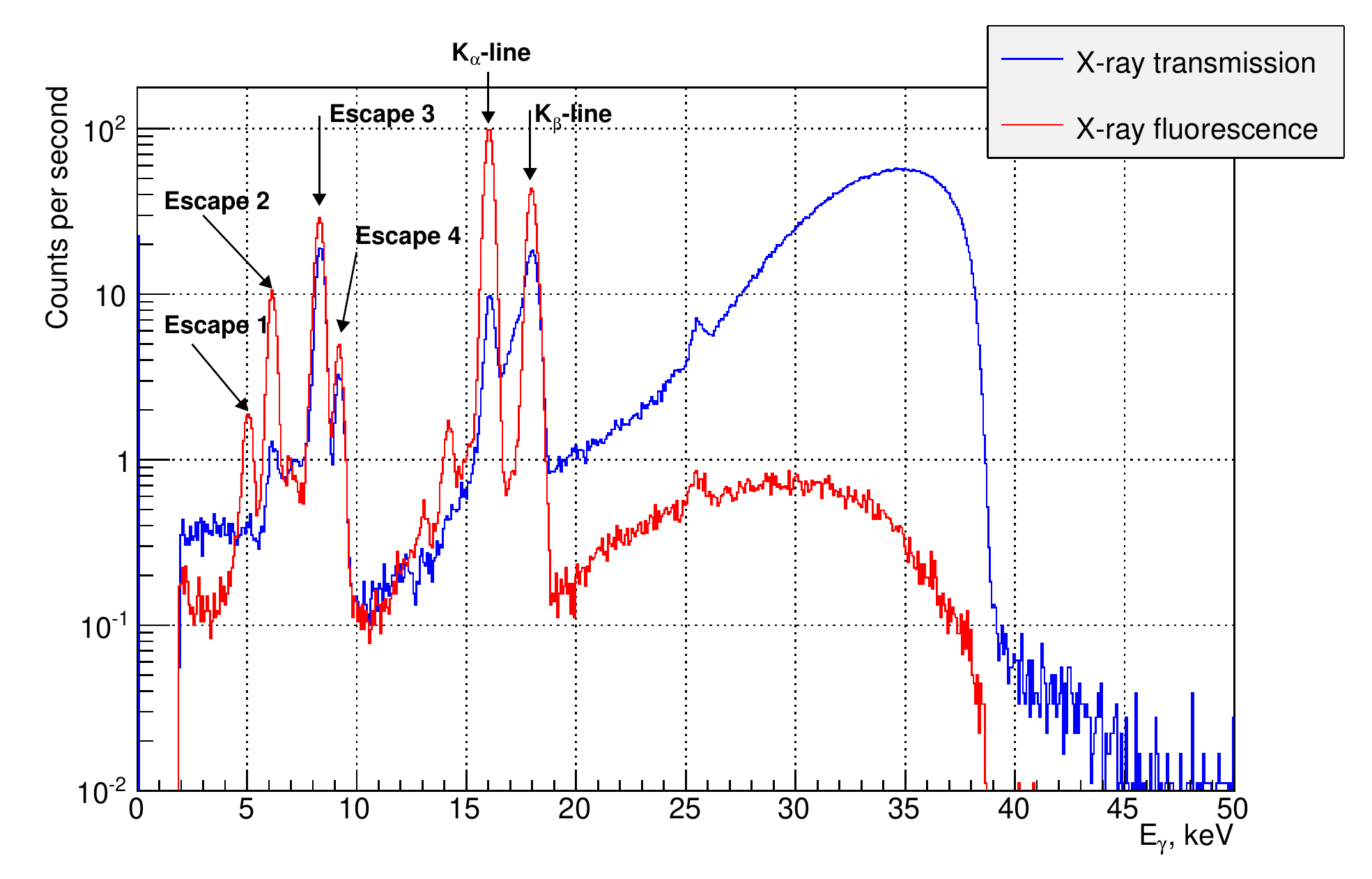}}
\caption{\footnotesize X-ray spectra from Zr in ``transmission'' (blue line) and ``reflection'' (red line) geometries. 
\label{Zr_comp}}
\end{figure}

In Fig.~\ref{Zr_comp} the spectra of X-rays passing through the zirconium foil (``transmission'' geometry) and reflected from the same foil (``reflection'' geometry) are shown. In both cases, the thickness of the zirconium foil was 0.3 mm, the working voltage was 45 kV, and the current of the tube was 100 $\mu$A. In these spectra not only peaks from the $K_{\alpha}$ and $K_{\beta}$ lines of the characteristic emission of zirconium are visible but also peaks (marked as Escape 1, Escape 2, Escape 3, and Escape 4) formed upon excitation of germanium atoms by primary photons of the zirconium characteristic emission with subsequent exiting of the germanium characteristic X-ray photons from the detector. The four additional peaks are due to the fact that a germanium atom can re-emit an absorbed photon both as $K_{\alpha}$ ($E^{Ge}_ {\alpha}\approx 9.9$ keV) and $K_{\beta}$ ($E^{Ge}_{\beta}\approx$ 11 keV) lines.

Although there are several lines in every area of energy $E^{Zr}_{\alpha}$ and $E^{Zr}_{\beta}$ (as seen from Table~\ref{MetalsKLines}), because the resolution of the detector is about 0.1 keV, the peaks with close energy are not distinguished and are merged together.

\begin{figure}[h!]
\center{\includegraphics[width=\linewidth]{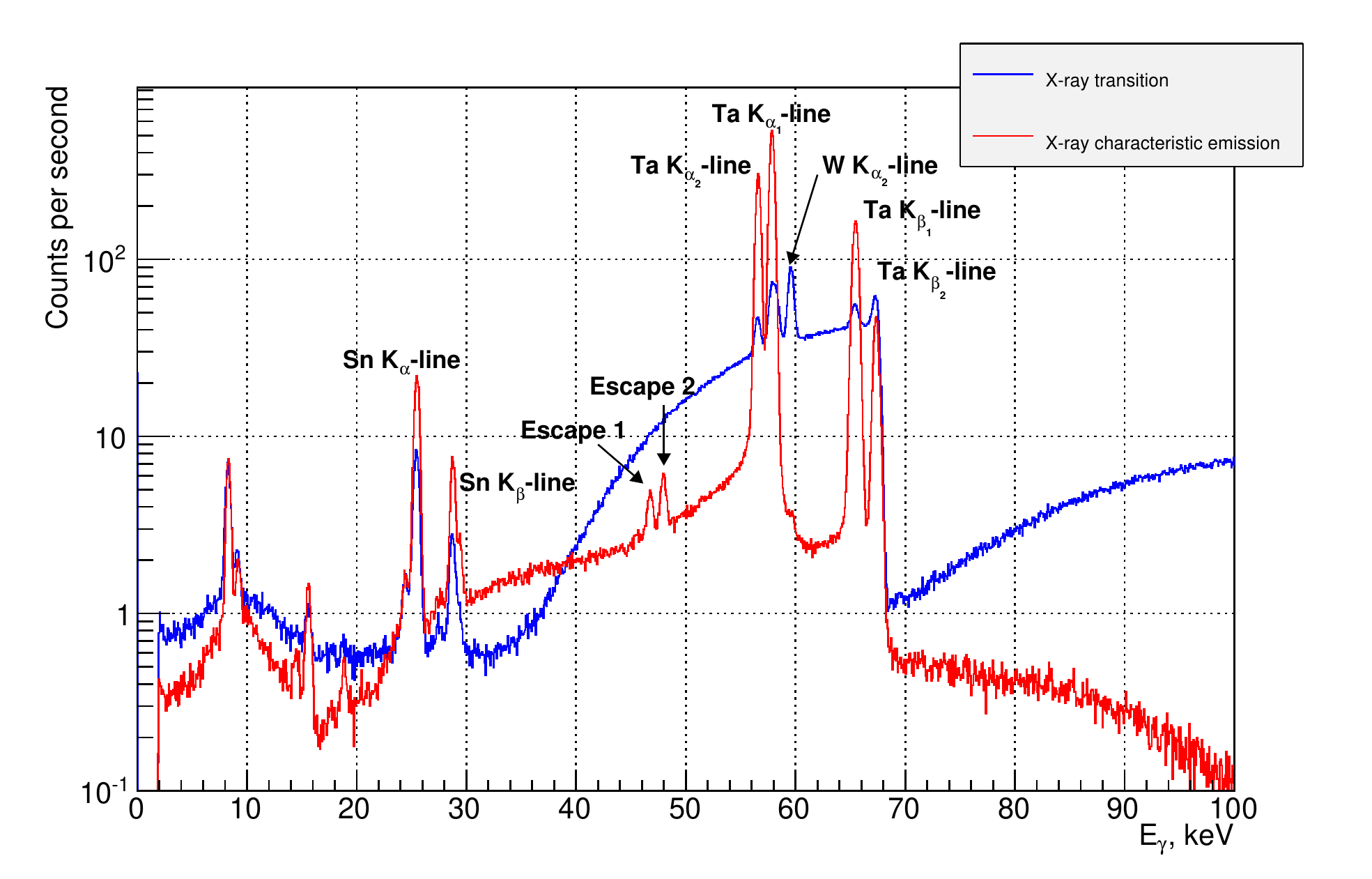}}
\caption{\footnotesize X-ray spectra from Ta in ``transmission'' (blue line) and ``reflection'' (red line) geometries.
\label{Ta_comp}}
\end{figure}

A similar comparison of the characteristic emission spectra obtained in ``transmission'' geometry and in ``reflection'' geometry is shown in Fig.~\ref{Ta_comp} for a 0.3 mm thick tantalum foil. In this case the operating voltage and the current of the X-ray tube were 140 kV and \mbox{30 $\mu$A}, respectively. In the presented spectra peaks from the $K_{\alpha_1}$, $K_{\alpha_2}$, $K_{\beta_1}$, and $K_{\beta_2}$ lines of the characteristic emission of tantalum are clearly distinguished. Furthermore, in the ``transmission'' spectrum a peak from the $K_{\alpha_1}$ line of the characteristic emission of tungsten appears. It should be expected since the X-ray tube anode is covered with tungsten. One can also notice the presence in the spectra of four additional peaks. Two of these peaks (marked as Escape 1 and Escape 2 in Fig.~\ref {Ta_comp}) correspond to the escape of characteristic X-ray photons of germanium from the detector (as a result of interaction with primary characteristic photons of tantalum). The other two (marked as Sn $K_{\alpha}$-line and Sn $K_{\beta}$-line in Fig.~\ref{Ta_comp}) are lines of characteristic emission of tin contained in duralumin, of which the bearing part of the internal curtain dividing the safety housing is made (see Fig.~\ref{Setup_2}).

From the figures it is clear that peaks corresponding to characteristic lines have a greater intensity for the measurements in the ``reflection'' geometry than similar peaks for the measurements in ``transmission'' geometry. For studies of the energy resolution and calibration of Timepix detectors it is fundamentally important to record spectra with well-defined lines of characteristic emission on continuous background. Therefore, for the calibration of Timepix detectors the setup with ``reflection'' geometry was chosen. This will be discussed in more details in the next section.

\begin{figure}[h!]
\center{\includegraphics[width=\linewidth]{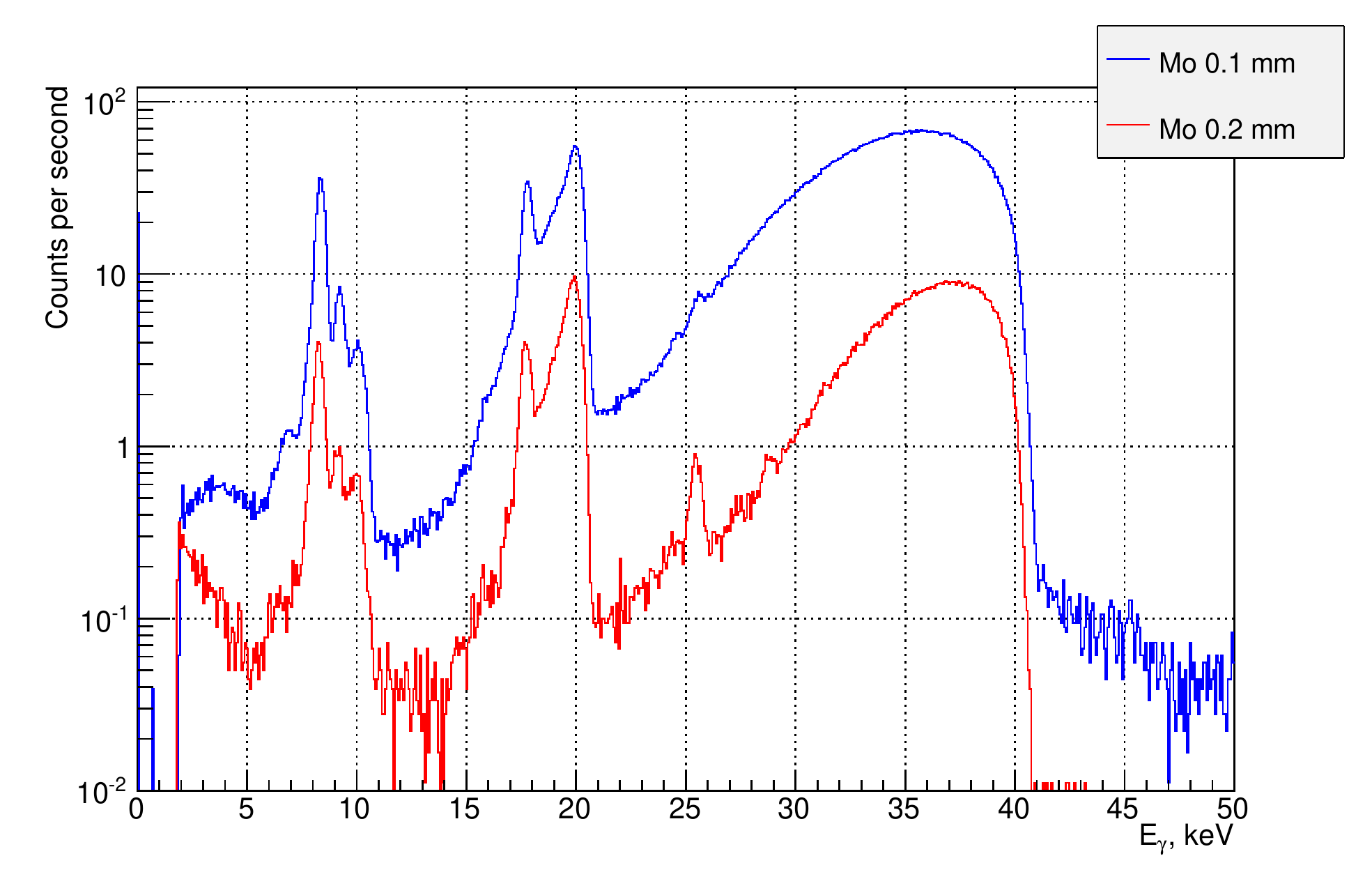}}
\caption{\footnotesize X-ray spectra in ``transmission'' geometry from a Mo foil of 0.1 mm (blue) and 0.2 mm (red) thickness.
\label{Mo_diff}}
\end{figure}

In Fig.~\ref{Mo_diff} a comparison of the spectra of X-rays passing through the molybdenum foils of 0.1mm and 0.2mm thickness is shown. In both cases the working voltage and the current were 45 kV and 100 $\mu$A, respectively. As one can see, the change in the target thickness only affects the intensity of the radiation, but not the form of the spectrum.

\clearpage

\section{Energy resolution and calibration of Timepix detectors}

\subsection{Timepix pixel detectors}

\begin{figure}[h]
\begin{minipage}[h]{0.5\linewidth} 
\center{\includegraphics[width=1.0\linewidth]{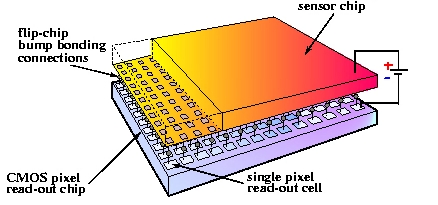} \\ a)} 
\end{minipage}
\hfill 
\begin{minipage}[h]{0.5\linewidth}
\center{\includegraphics[width=1.0\linewidth]{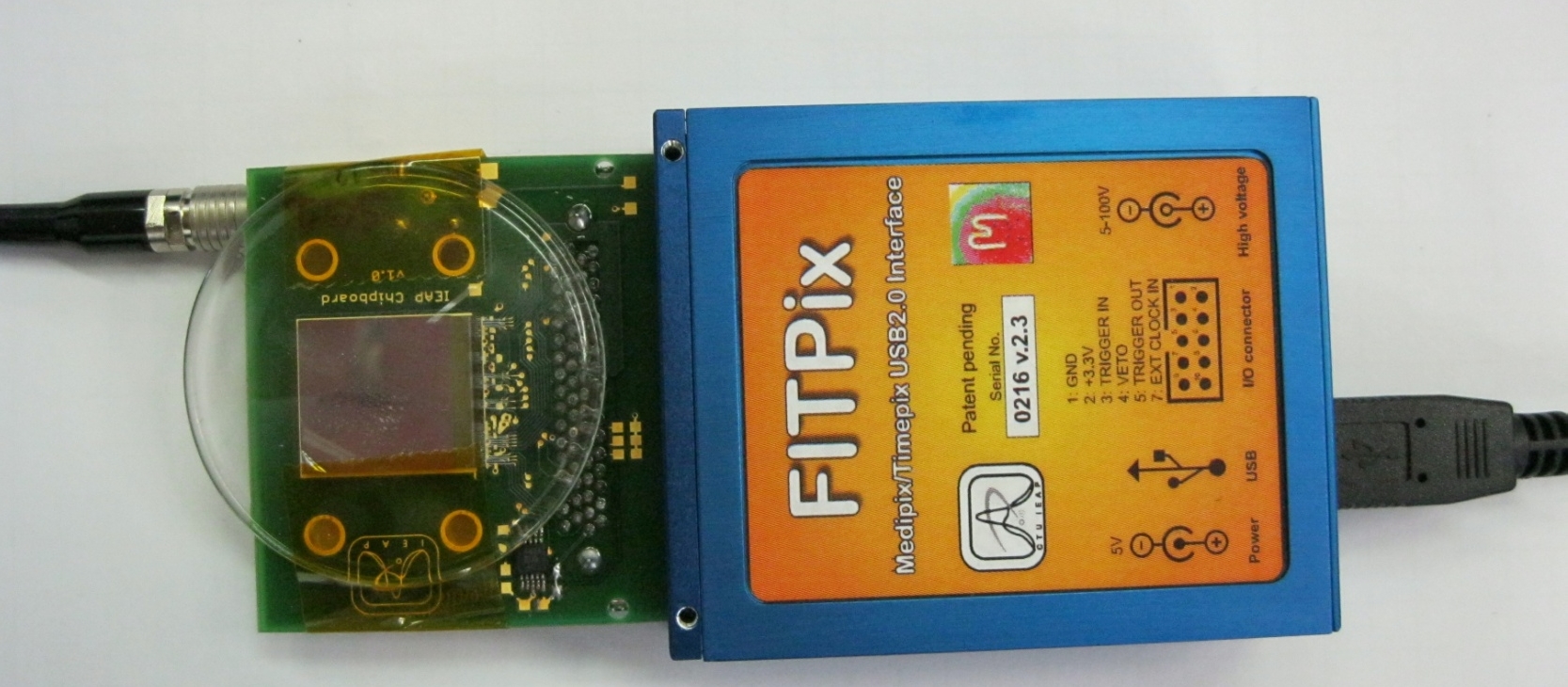} \\ b)}
\end{minipage}
\caption{\footnotesize a) Structure of a Timepix detector; b) Timepix pixel detector with USB interface FITPix.} \label{TimepixDevice}
\end{figure}

The Timepix readout chips were developed in CERN by Medipix collaboration with support from EUDET. Detectors based on these chips can be used simultaneously for tracking and measuring of particle energy. The structure of hybrid pixel detectors based on the Timepix chip is shown in Fig.~\ref{TimepixDevice}a. The detector consists of two main parts - the sensor (in this work of gallium arsenide doped with chromium GaAs:Cr), in which a charged particle leaves a ``trace'' as a certain number of free charge carriers proportional to the deposited energy, and the Timepix chip, wherein the charge formed in the sensor is amplified, analyzed, digitized, and transferred to a computer. The thickness of GaAs:Cr sensor is determined by the capabilities of its production technology and can vary at the moment from 0.2 mm to 1 mm. The sensor dimensions are defined by the size of the Timepix chip and are equal to 14.2$\times$14.2 mm$^2$. Also, the geometry of the sense electrodes on the sensor surface is defined by the geometry of the Timepix readout chip. The semiconductor sensor and the readout chip and connected with the ``flip-chip'' technology. The Timepix readout chip consists of 65,536 independent channels (pixels) assembled in a matrix of 256$\times$256 elements. The single pixel size is 55$\times$55 m$^2$. Each pixel has an amplifier, a discriminator with adjustable threshold, and a counter. The first results of the studies of hybrid pixel detectors with GaAs:Cr sensors and the Timepix readout chips are given in~\cite{medipixJINR}. In Fig.~\ref{TimepixDevice}b one of the first GaAs:Cr Timepix detectors assembled in 2008 for the LNP is depicted, along with the FITPix~\cite{fitpix} USB interface that connects the detector to a PC.

\begin{figure}[h]
\center{\includegraphics[width=0.5\linewidth]{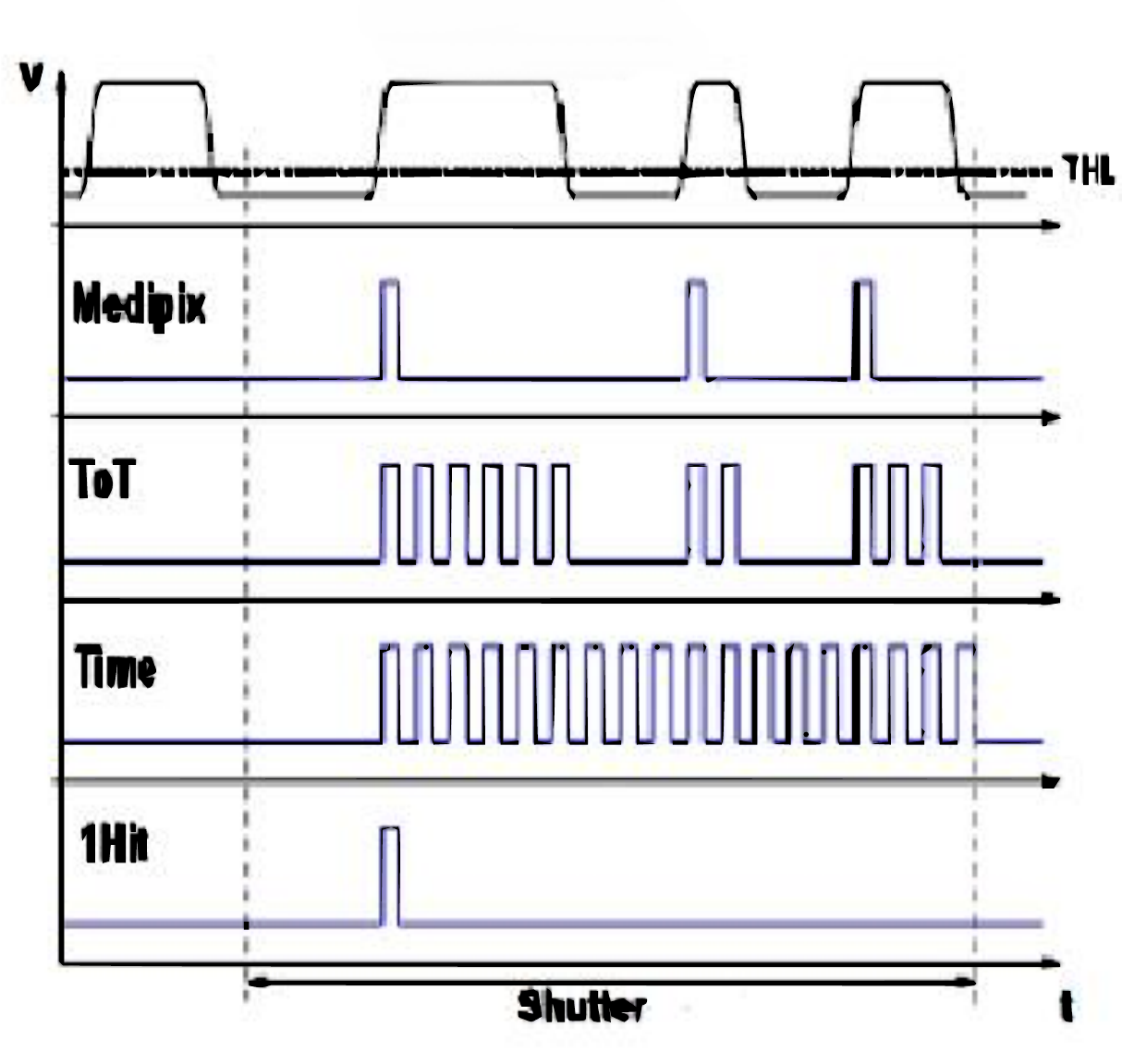}}
\caption{\footnotesize Modes of operation of Timepix detector. $THL$ -- threshold value.\label{Timepixmode}}
\end{figure}

The Timepix chip can operate in one of four modes (Fig.~\ref{Timepixmode}):
\begin{itemize}
\item Medipix mode: the chip counts how many times during the open shutter the signal exceeds a certain threshold;
\item Time-over-Threshold (ToT) mode: the chip measures for how long the signal stays above the threshold;
\item Time-of-arrival (ToA) Mode: the chip measures time from the moment when the signal crosses the threshold until the shutter is closed;
\item OneHit mode: the chip checks if the signal exceeds the threshold at least once during the open shutter.
\end{itemize}

To measure energy deposited in the sensor by a particle the $ToT$ mode of the Timepix chip is used. In this mode, when the signal in a pixel crosses some threshold level $THL$ the pixel counter starts counting pulses from the oscillator. It stops when the amplitude of the input signal falls below the threshold. The number of counted pulses depends on the amount of absorbed energy. The oscillator frequency can be set in the range of 10-100 MHz.

\begin{figure}[h]
\center{\includegraphics[width=0.8\linewidth]{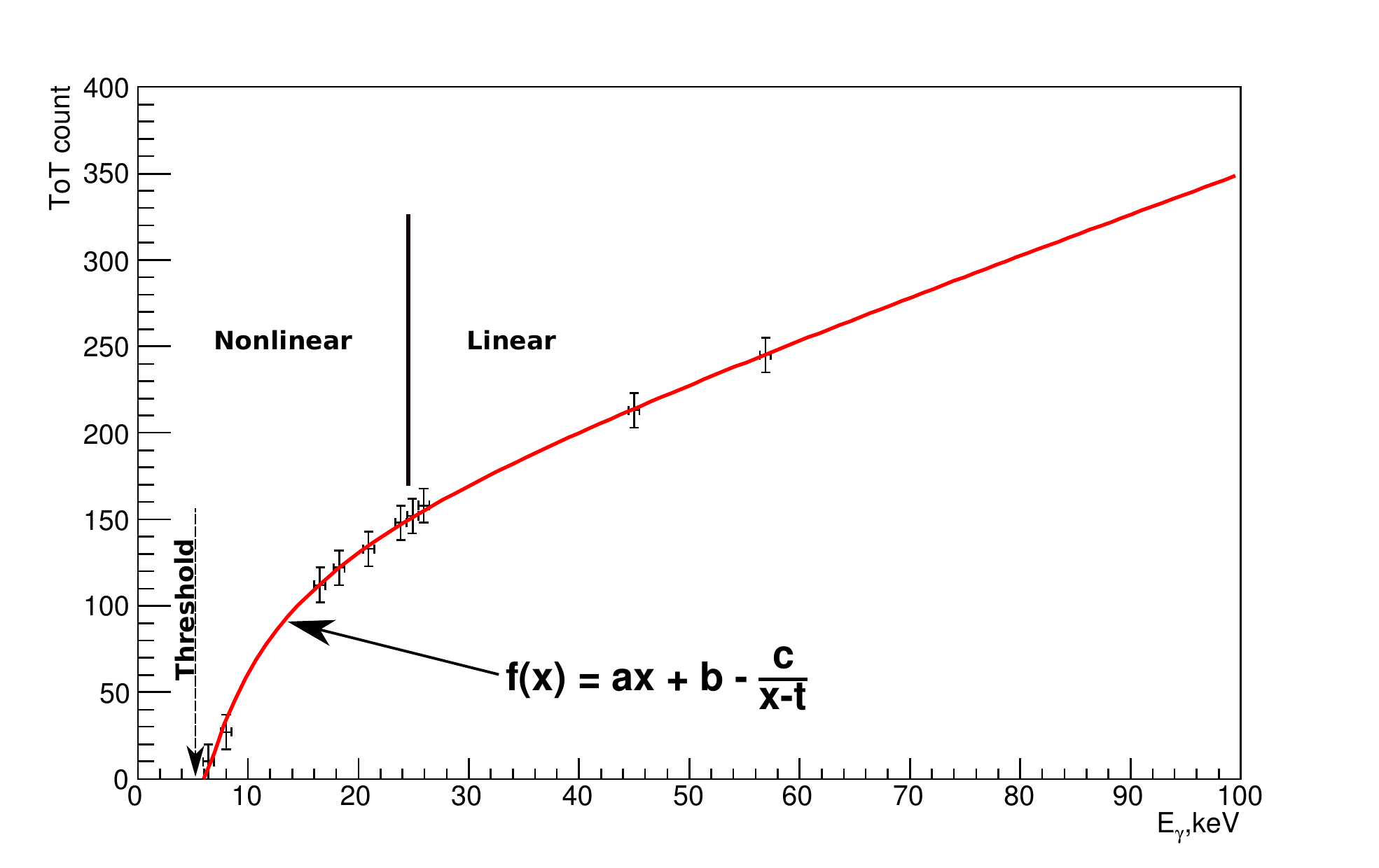}}
\caption{\footnotesize General shape of the calibration curve (\ref{eq.ToT(E)}). \label{shemaToT(E)}}
\end{figure}

To determine the relationship between the value of $ToT$ counts and the value of deposited energy $E$ an energy calibration of the detector is required. The $ToT$-$E$ relationship can be approximated by the following function (its general view is shown in Fig.~\ref{shemaToT(E)}):
\begin{equation}                
 \label{eq.ToT(E)}
 \begin{aligned}                
  ToT(E)=a \cdot E+b+\frac{c}{E-t}.
 \end{aligned}
\end{equation}
Its reverse function is the following:
\begin{equation}                
 \label{eq.E(ToT)}
 \begin{aligned}               
  E(ToT)=\frac{a \cdot t + ToT - b + \sqrt{(b + a \cdot t - ToT})^2 + 4 \cdot a \cdot c}{2a}.
 \end{aligned}
\end{equation}
Empirical formulas (\ref{eq.ToT(E)}) and (\ref{eq.E(ToT)}) were first proposed in~\cite{calfunction}.

Due to non-uniformity of the sensor material and the spread of parameters for individual channels of the Timepix readout chip, the coefficients $a$, $b$, $c$ and $t$ of the calibration curve will be different for different pixels. That means that procedure for the energy calibration must be done on per-pixel basis. The per-pixel calibration, along with the generalized calibration, when the coefficients $a$, $b$, $c$ and $t$ are set equal for the entire pixel matrix, will be discussed in detail in the next section.


\subsection{Procedure for measurement of energy resolution and calibration of Timepix detector}

This section describes the procedure for the generalized and per-pixel calibrations of Timepix detectors with GaAs:Cr sensors. The procedure also includes measurement of the energy resolution. While the basic idea of this procedure is simple and has been used successfully on the detectors with Si as a sensing layer~\cite{TimepixCalib}, the specific properties of GaAs:Cr as a material and particularities of detector prototypes based on it require some changes in the calibration procedure.

In this work, we used a hybrid Timepix detector with a GaAs:Cr sensor of 500 $\mu$m thickness. The bias voltage applied to the sensor was $V_{bias}=-500$ V. To exclude the effects of temperature the sensor was thermally stabilized (at 18$^{\circ}$C) with the use of a Peltier element with a heat output of up to 5 watts.

Calibration of the pixel detector is carried out in two stages. The first stage is the pixels threshold equalization. For this purpose the Pixelman~\cite{Pixelman} software package is used. The threshold level $THL$ is set the same for all pixels of the readout chip. But the electrical properties of each pixel are somewhat different due to technological features that lead to each pixel having its own noise level. Therefore, an additional corrective bias voltage is introduced into the amplifier channel of each pixel in order to equalize noise levels of all pixels at the set threshold level $THL$.

The second step is the establishment of a one-to-one correspondence between the time length of the signal in $ToT$ counts and the absorbed energy $E$. The procedure for obtaining $E(ToT)$ dependence is called ToT-calibration. It can be done individually for each pixel (per-pixel calibration) and for the entire matrix as a whole (generalized calibration). Generally ToT-calibration comprises of the following steps:
\begin{itemize}
\item measurements of X-ray spectra of known radioactive $\gamma$-sources and spectra of characteristic emission of selected materials in units $ToT$;
\item identification of peaks in the recorded ToT-spectra with corresponding X-ray emission lines;
\item fitting of the peaks and constructing of a $ToT(E)$ relation graph;
\item fitting of the obtained $ToT(E)$ graph with the function (\ref{eq.ToT(E)}) and calculation of the inverse relationship $E(ToT)$ according to (\ref{eq.E(ToT)}).
\end{itemize}

The spectra of characteristic emission of Zr, Mo, Rh, Cd, In, Sn, and Ta were chosen as the benchmark spectra for the energy calibration (their K-lines are listed in Table~\ref{MetalsKLines}). This choice of foils allows to calibrate detectors in the energy range of [15,100] keV and takes into account the presence of two regions in the calibration curve. For the linear region above $\sim$25 keV two reference energy points are needed while in the nonlinear region below 25 keV more control points are require to define the shape of the calibration curve with sufficient accuracy. 

As mentioned above, the characteristic emission spectra used in the energy calibration were measured in the ``reflection'' geometry of the experimental layout (shown in Fig.~\ref{Setup_2}). The microfocus X-ray tube RAP-150MN was used as the X-ray source. Recording of ToT spectra was done with the Pixelman software package. Data processing was carried out in the ROOT data analysis framework. By adjusting the position of the Timepix detector relative to the X-ray tube  uniform illumination of the detector matrix was achieved. For all measurements the clock frequency of the Timepix pulse generator was set to $f=10$ MHz, the threshold to $THL=380$, the duration of the open shutter (time frame) to $T_f=0.05$ s.

It is well known that the spectrum of characteristic X-ray emission of a chemical element consists of a set of closely spaced $K_{\alpha}$, $K_{\beta}$, $L_{\alpha}$ spectral lines. The hybrid Timepix detectors do not have good enough energy resolution to separate them, therefore in the spectra measured with such detectors these lines merge into a single peak. In order to determine the energy corresponding to the maximum of a peak in a spectrum recorded by a Timepix detector, a simple Monte Carlo simulation was conducted that takes into account the detector resolution and efficiency. The Monte Carlo simulation consisted of the following steps:
\begin{itemize}
\item in the spectrum of characteristic X-ray emission of a particular element obtained with the Canberra detector, a region containing the element's $K_{\alpha}$ and $K_{\beta}$ spectral lines is selected;
\item for the $i$-th bin of the selected region with the mean energy $\mu_i$ a Gaussian $G_i=(N_i,\mu_i,\sigma_i)$ is generated, where $ N_i = S_i \cdot m $ - the number of generated events with $ S_i $ being the number of photons in the $i$-th bin of the original spectrum and $m$ is a constant at the order of 10$^3$ to even out the statistical fluctuations, $ \sigma_i = \sigma_ {E} ^ {MC} \cdot \mu_i $ with $ \sigma_ {E} ^ {MC} $ being the energy resolution of the detector at the given energy assumed for the model;
\item the generated Gaussians $ G_i $ for all bins in the selected area are summed over:
$$S_{model}=\frac{1}{m} \cdot \sum\limits_{i} G_i.$$
\item the photon detection efficiency of the detector is taken into account in accordance with the Beer–Lambert–Bouguer law:
$$\varepsilon(E)=1 - \exp \left(-\frac{\mu}{\rho}(E) \cdot \rho \cdot d \right),$$
where $\varepsilon$ -- probability of photon absorption by material of thickness $ d $ and density $ \rho $, $ \frac {\mu} {\rho} $ - the material's X-ray mass attenuation coefficient depending on the photon energy.
\end{itemize}

Since before the calibration, the energy resolution of the Timepix detector $ \sigma_{E}^{MC} $ is unknown, for the first iteration of the simulation the values of the relative half-widths of the peaks in the ToT spectra summed over all pixels are used. These values of $ \sigma_ {E}^{MC}$ are proved to be 1.5-2 times larger than the values of $ \sigma_{E}^{meas} $ in the energy spectra recorded by the Timepix detector. Accordingly, the values of $ \sigma_{E}^{MC} $ are adjusted for the second iteration of the simulation. This iterative process can be carried out until the difference between $ \sigma_{E}^{MC} $ and $ \sigma_{E}^{meas} $ becomes small. Typically, two iterations are sufficient.

\begin{figure}[htb]
\begin{minipage}[h]{0.5\linewidth} 
\center{\includegraphics[width=1.0\linewidth]{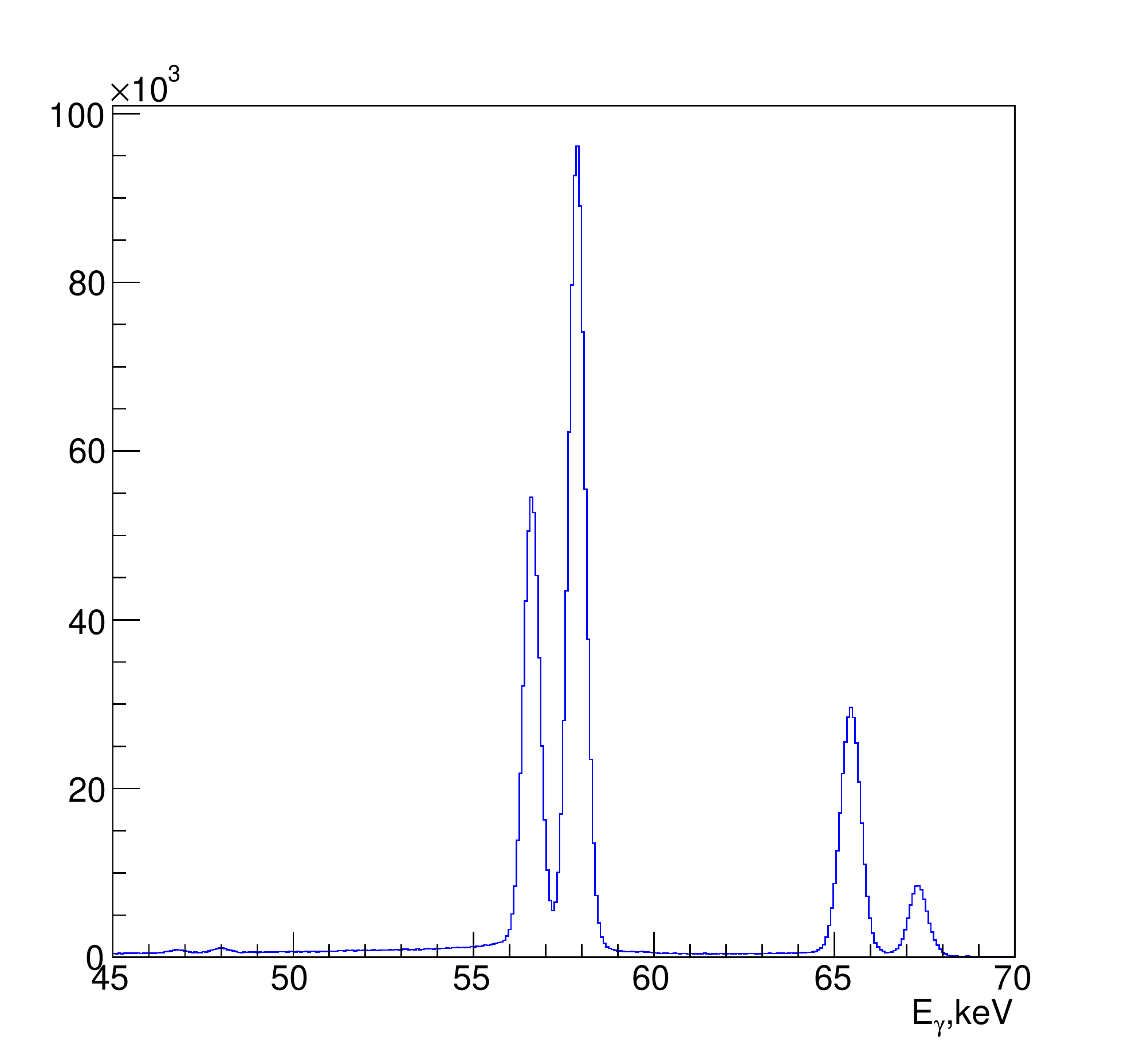} \\ a)} 
\end{minipage}
\hfill 
\begin{minipage}[h]{0.5\linewidth}
\center{\includegraphics[width=1.0\linewidth]{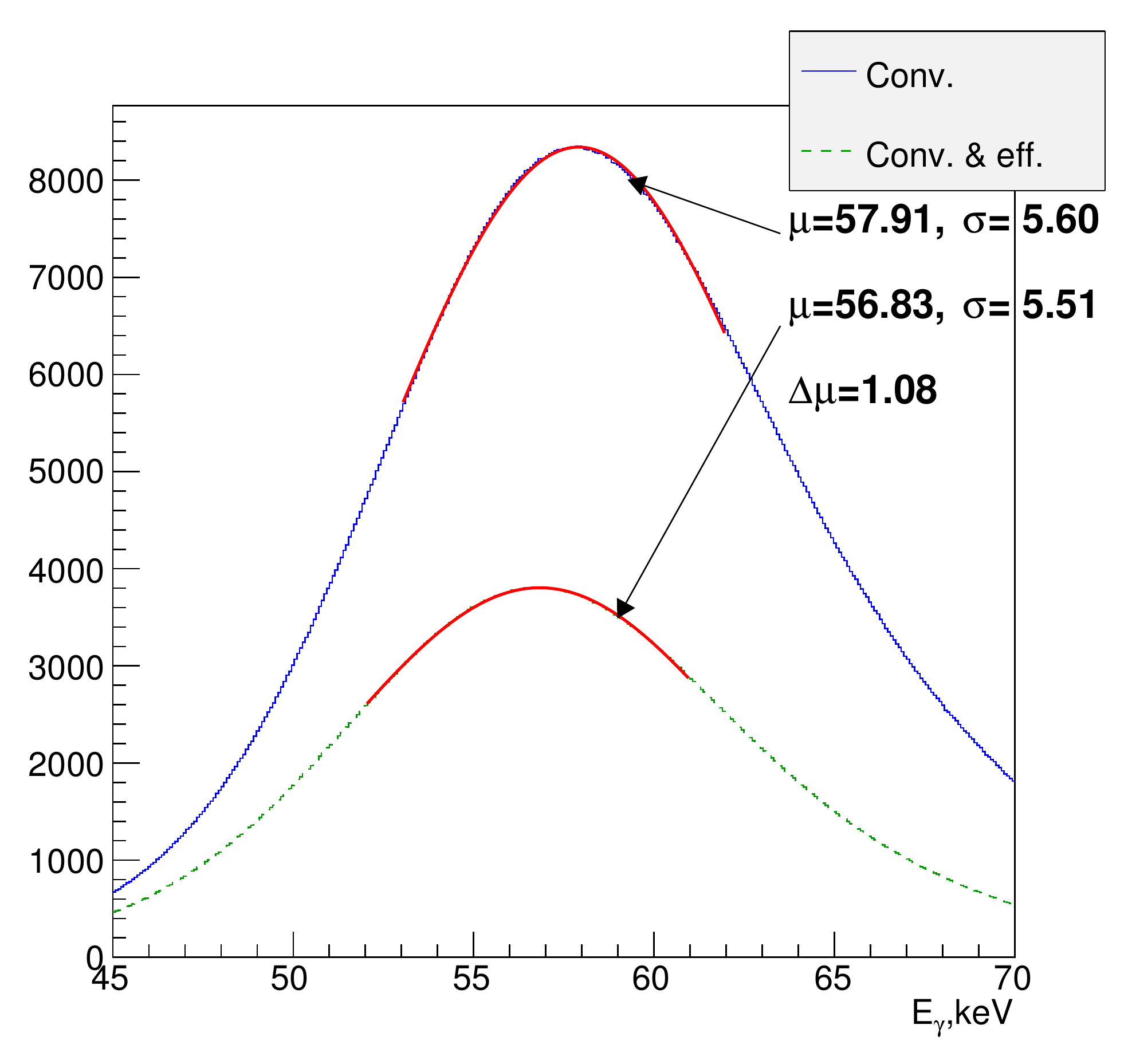} \\ b)}
\end{minipage}
\caption{\footnotesize Spectrum of Ta characteristic X-ray emission in [45;70] keV range: a) measured with Canberra detector; b) modeled by convolution of the Canberra spectrum with the Timepix resolution (blue) plus absorption efficiency in GaAs (green).} \label{Convolution}
\end{figure}

In Fig.~\ref{Convolution} the Ta characteristic emission spectrum measured with the Canberra detector and the spectrum modeled for a GaAs:Cr Timepix detector according to the description above are compared. Once the efficiency of a 500 $\mu$m thick GaAs sensor is taken into account in the modeling, not only the shape of the spectrum is changed (see Fig.~\ref{Convolution}b), but also its maximum is shifted ($\Delta \mu = 1.08 $ keV). Thus the described Monte Carlo simulation allows to achieve more accurate determination of positions of the reference peak energies and improve the accuracy of the energy calibration of the Timepix detector.

\subsection{Generalized calibration of Timepix detector}

\begin{figure}[hbt]
\center{\includegraphics[width=\linewidth]{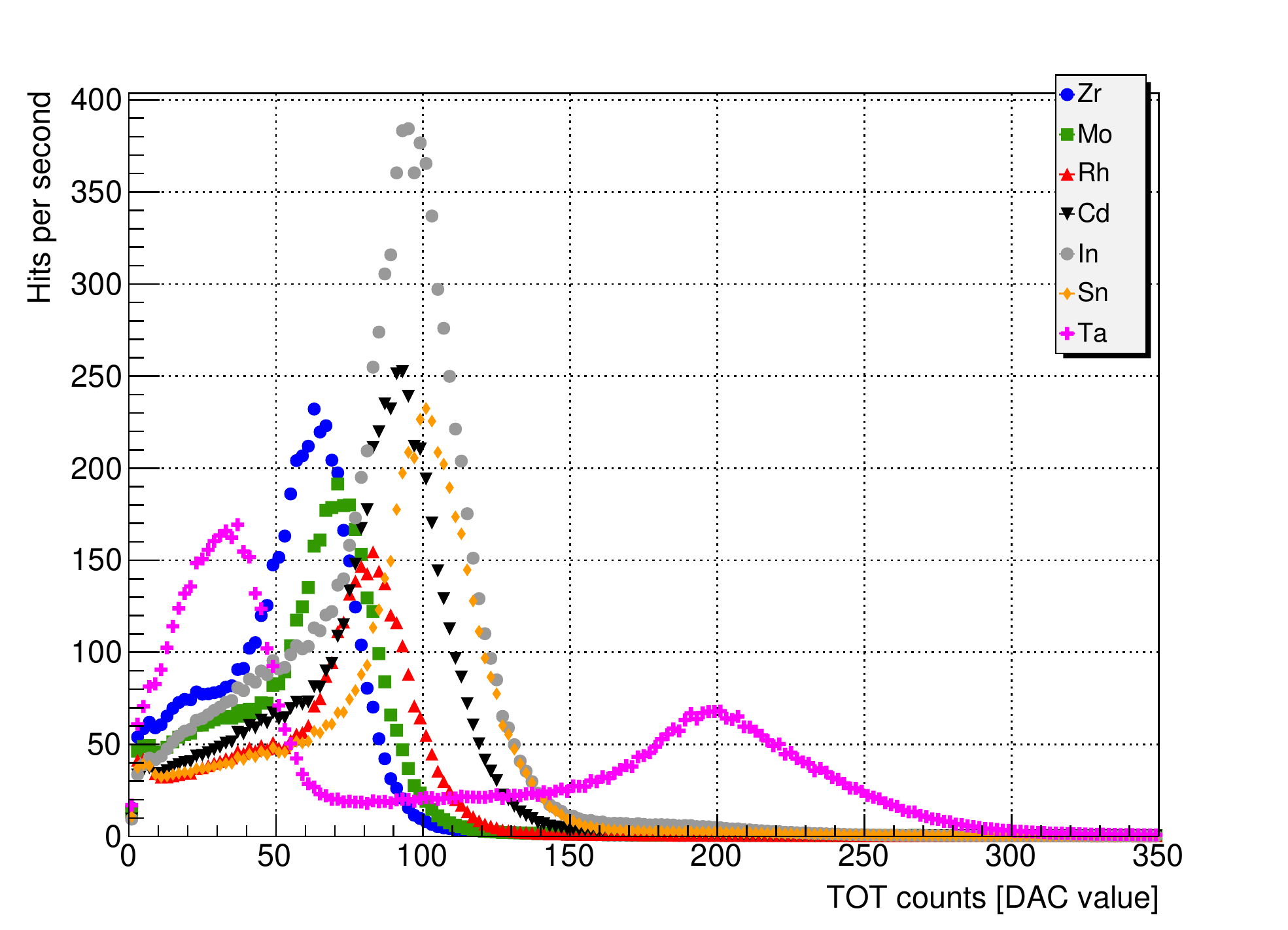}}
\caption{\footnotesize Summed ToT spectra of different metals used for generalized calibration of Timepix detector.}
\label{GenToTSpectraAll}
\end{figure}

Initial data for the generalized calibration of Timepix detectors are the total TOT spectra obtained by summing statistics from all pixels. Examples of such spectra are shown in Fig.~\ref{GenToTSpectraAll}. This approach allows quickly calibrate the detector, since the time of data collection in this case is a few orders of magnitude smaller than the time of per-pixel calibration that needs considerable number of hits in every pixel. But the accuracy of generalized calibration is worse, because it does not take into account differences between the pixels caused by non-uniformity of the sensor material and quality of the detector assembly.

During the data collecting in order to eliminate the influence of the charge sharing effect~\cite{aladdin4} only single-pixel events were selected when absorbed photons create hits only in one pixel. The charge sharing appears when a single particle generates hits in several adjacent pixels. This is because the electron-hole cloud created in the sensor material by the absorbed or passing particle increases in size as it drifts to the electrodes, and eventually the charge can be shared by several pixels forming a cluster.

\begin{figure}[htb]
\begin{minipage}[h]{0.5\linewidth} 
\center{\includegraphics[width=1.0\linewidth]{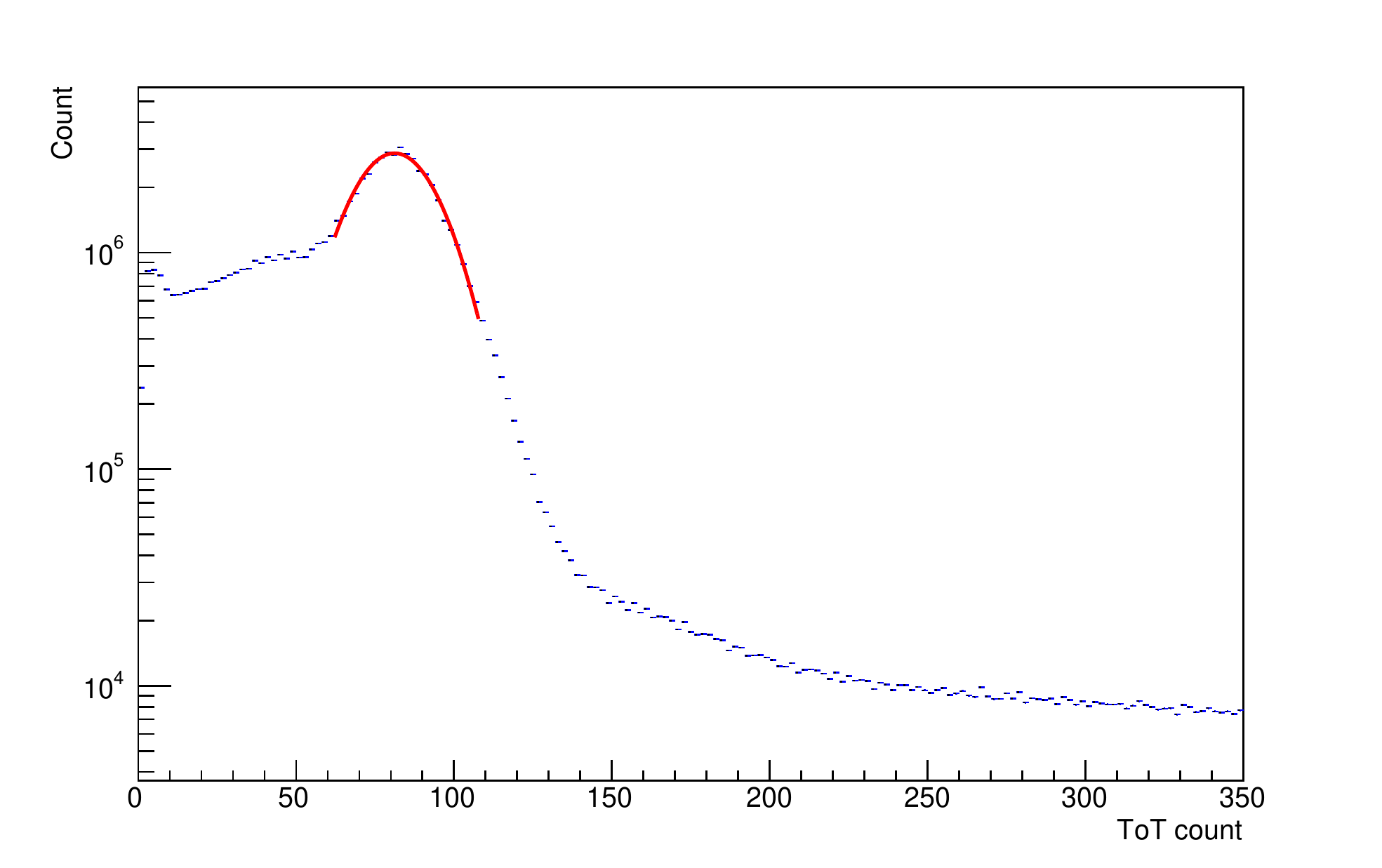} \\ a)} 
\end{minipage}
\hfill 
\begin{minipage}[h]{0.5\linewidth}
\center{\includegraphics[width=1.0\linewidth]{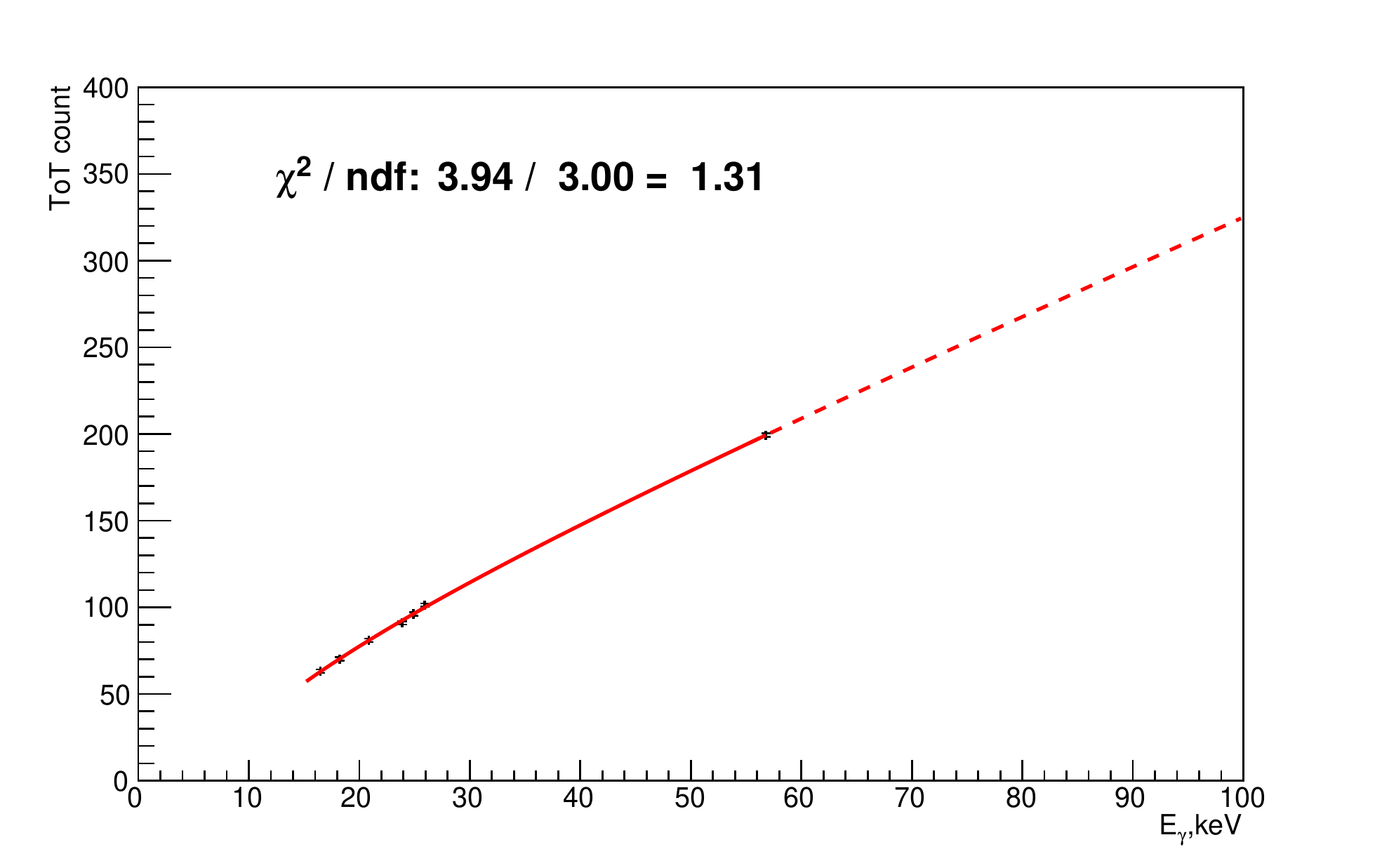} \\ b)}
\end{minipage}
\caption{\footnotesize a) Gaussian fit of the summed over pixels ToT spectrum of characteristic X-ray emission from Rh;
b) Generalized calibration curve, fitted over seven data points:  Zr, Mo, Rh, Cd, In, Sn, Ta.} \label{GenClb}
\end{figure}

An important step of the calibration is the fitting of the reference spectral peaks to extract quantitative information about their positions and widths. The peaks in the spectra have the normal distribution shape, so they were fitted using a Gaussian function with parameters properly chosen for each individual peak of characteristic emission, as is shown in Fig.~\ref{GenClb}a for the spectrum of rhodium.

Knowing the ToT values of the peak positions and the mean energies of their corresponding characteristic emission lines found using the Monte Carlo simulation, we can construct a graph of $ ToT(E) $. The fit to this graph is performed in two steps. Firstly, in the energy range $[25,100]$ keV a straight line $ a \cdot E + b $ is used as the fitting function and the values of the parameters $a$ and $b$ are determined as well as their errors. These errors define the range of variation for the parameters $a$ and $b$ in the second step when all data points in the energy range $[15,100]$ keV are fitted with the function (\ref{eq.ToT(E)}). As a result, all four parameters $ a $, $ b $, $ c $, $ t $ are determined. The result of the generalized calibration of the Timepix detector \mbox{E11-W0110} in the energy range $ [15,100] $ keV is shown in Fig.~\ref{GenClb}b with the obtained parameter values: $ a = 2.78 $, $ b = 57.25 $, $ c = 1020 $, $ t = -8.96 $.


\subsection{Per-pixel calibration of Timepix detector}

Since all pixels in a hybrid pixel detector give a somewhat different response to the passage of a particle with a certain energy, in order to have the best energy resolution of the detector the parameters of the calibration curve must be determined individually for each pixel. The basic idea of the calibration process is the same as in the generalized calibration, so only particular features of the per-pixel calibration and its automation will be discussed below.

In the per-pixel calibration ToT spectra obtained in each pixel are considered. Thus, a Timepix detector needs 65536 ToT spectra for a single energy point in the calibration curve to be processed and analyzed. To perform the fitting procedure on pixel spectra one needs to collect a considerable amount of statistics for each pixel. The majority of the measurements during the per-pixel calibration is carried out until about 1000 hits per pixel have been collected. The following algorithm is implemented for automation of the per-pixel calibration procedure:
\begin{itemize}
\item collecting of necessary statistics to build per-pixel ToT spectra of the characteristic emission of the selected element;
\item selecting in the spectra of each pixel a region containing a peak of known energy (the region is chosen individually for each spectrum);
\item fitting the selected region of the spectrum with a combined function.
\end{itemize}

Because the shape of the spectra from different elements varies from pixel to pixel, the limits on the fitting function parameters are selected for each ToT spectrum individually to ensure a better convergence of the fit. The range of the fitting function is defined as some neighborhood of the bin with the maximum value. The maximum likelihood method is used in the fitting, since it gives better results in the case of bins with low statistics. If a pixel is dead or is not working properly, the values of the fitting parameters for such a pixel are set to the result of the fit of the summed ToT spectrum.

This procedure is conducted for all the selected target materials. Once a sufficient number of data points is collected, fits to the calibration curve are performed for each pixel individually similar to the case of the generalized calibration. As a result, one receives 65536 calibration curves relating the value of ToT measured by a pixel in the detector and the energy deposited in this pixel.

The shapes of per-pixel spectra measured by the Timepix detector are asymmetric and differ from the normal distribution. It is especially noticeable in the nonlinear low energy part of the calibration curve (Fig.~\ref{shemaToT(E)}), where thermal noise of the sensor and noise in electronics are high. Fitting of a pixel spectrum using Gaussian gives systematically erroneous results, therefore per-pixel spectra are fitted with the combined function that is the sum of the quadratic polynomial (to take noise into account) and a Gaussian as follows:
\begin{equation}               
 \label{modelfunc}
  F=
  \begin{cases}
  K \cdot \exp \left(-\frac{(ToT-\mu)^2}{2\sigma^2} \right), &\text{if $ToT>\mu$;}\\
  K \cdot \exp \left(-\frac{(ToT-\mu)^2}{2\sigma^2} \right) + a \cdot ToT^2+b \cdot ToT+c,&\text{if $ToT<\mu$;}
  \end{cases}
\end{equation}
with additional requirement on the polynomial, $a \mu^2+b \mu+c=0$.

\begin{figure}[h]
\begin{minipage}[h]{0.5\linewidth} 
\center{\includegraphics[width=1.0\linewidth]{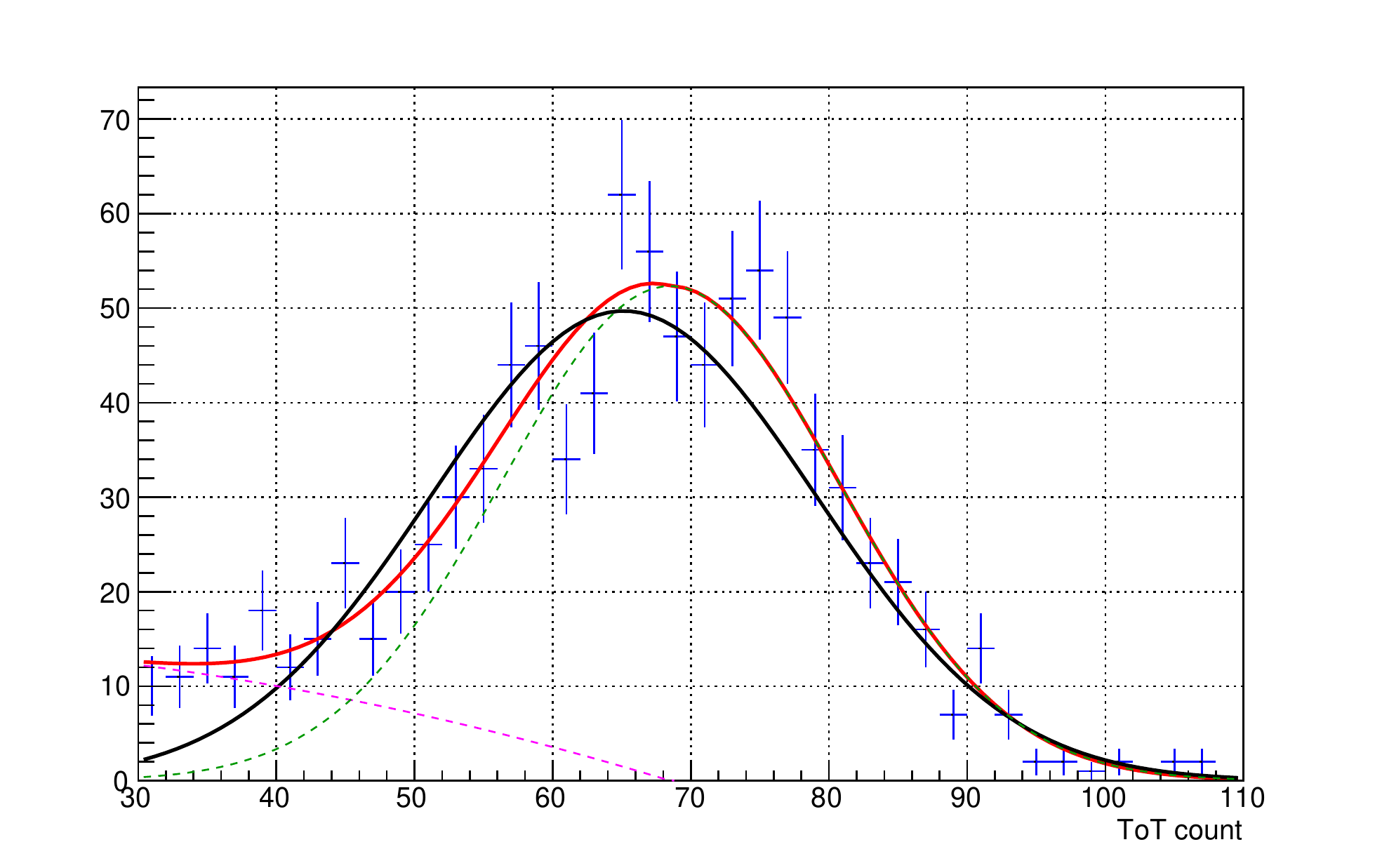} \\ a)} 
\end{minipage}
\hfill 
\begin{minipage}[h]{0.5\linewidth}
\center{\includegraphics[width=1.0\linewidth]{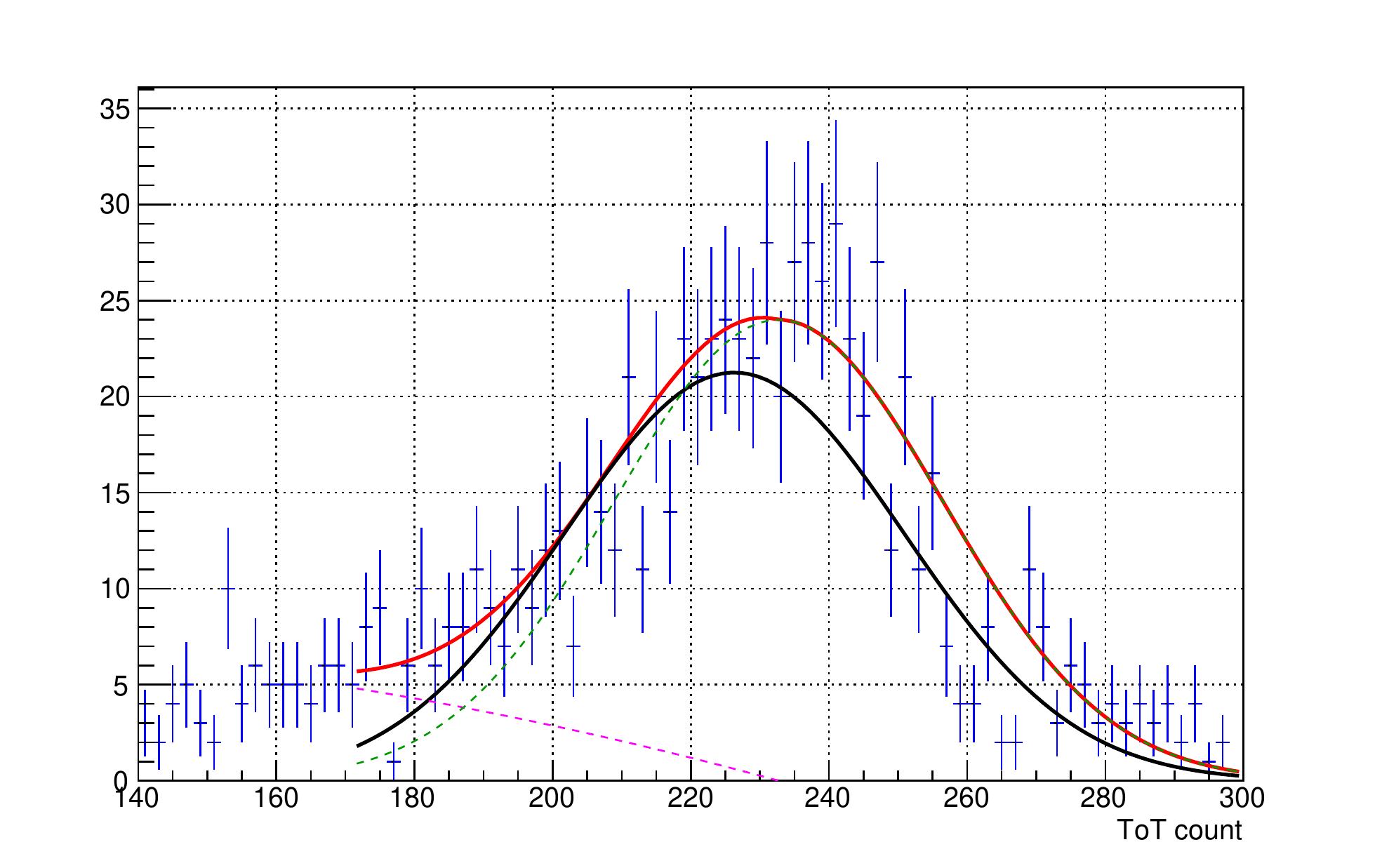} \\ b)}
\end{minipage}
\caption{\footnotesize Fitting with function (\ref{modelfunc}) a single pixel ToT spectra from Rh (a) and Ta (b).}
\label{PxlFit}
\end{figure}

In Fig.~\ref{PxlFit} the results of fitting of the ToT spectra of characteristic emission from rhodium and tantalum, registered by one of the pixels, are presented. One can clearly see differences between the pure Gaussian fit (black line) and the fit with the combined function of~(\ref{modelfunc}) (red line). The green and purple dotted lines represent contributions to the combined function from the Gaussian and the quadratic polynomial, respectively.

\begin{figure}[hbt]
\begin{minipage}[h]{0.5\linewidth} 
\center{\includegraphics[width=1.0\linewidth]{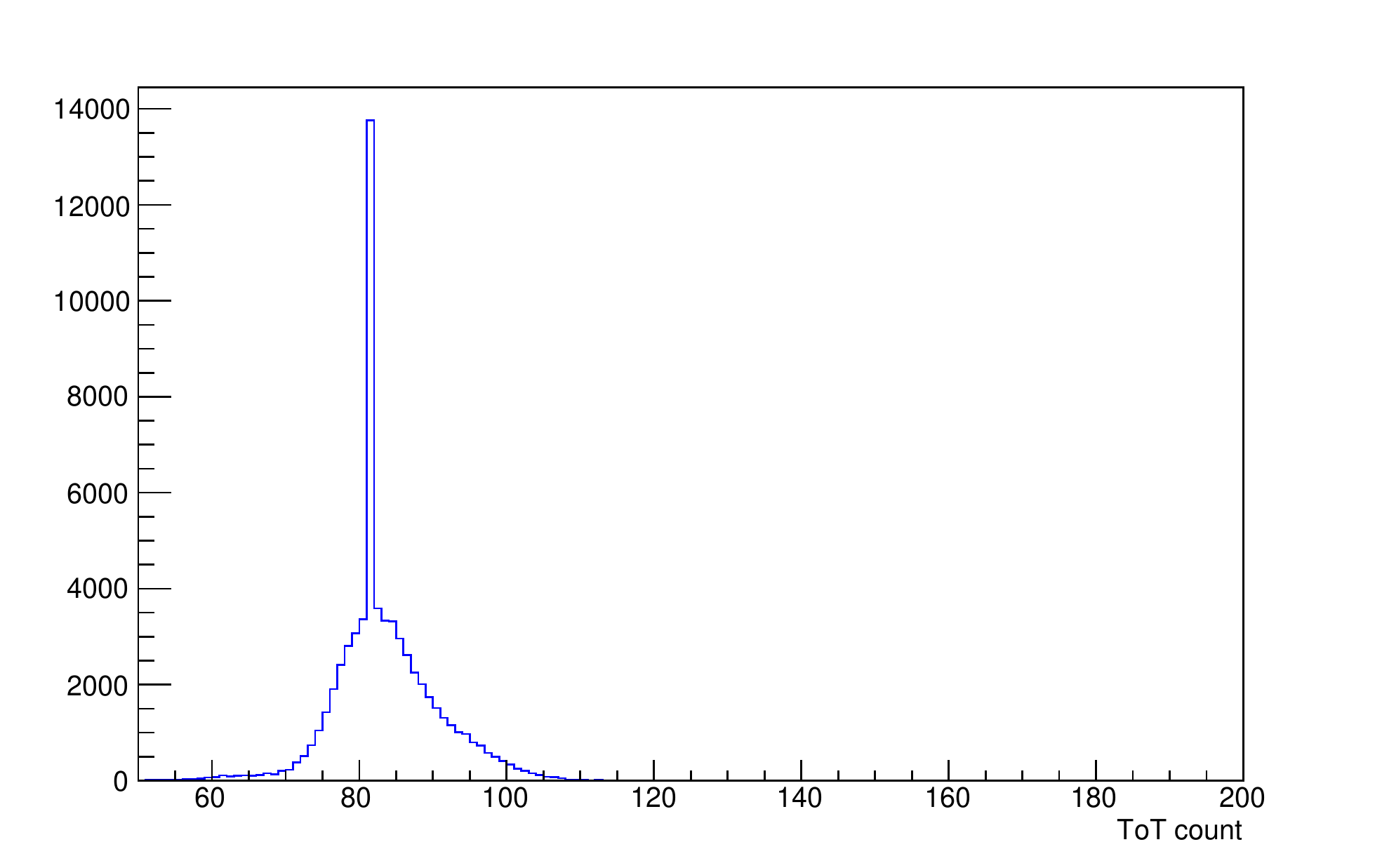} \\ a)} 
\end{minipage}
\hfill 
\begin{minipage}[h]{0.5\linewidth}
\center{\includegraphics[width=1.0\linewidth]{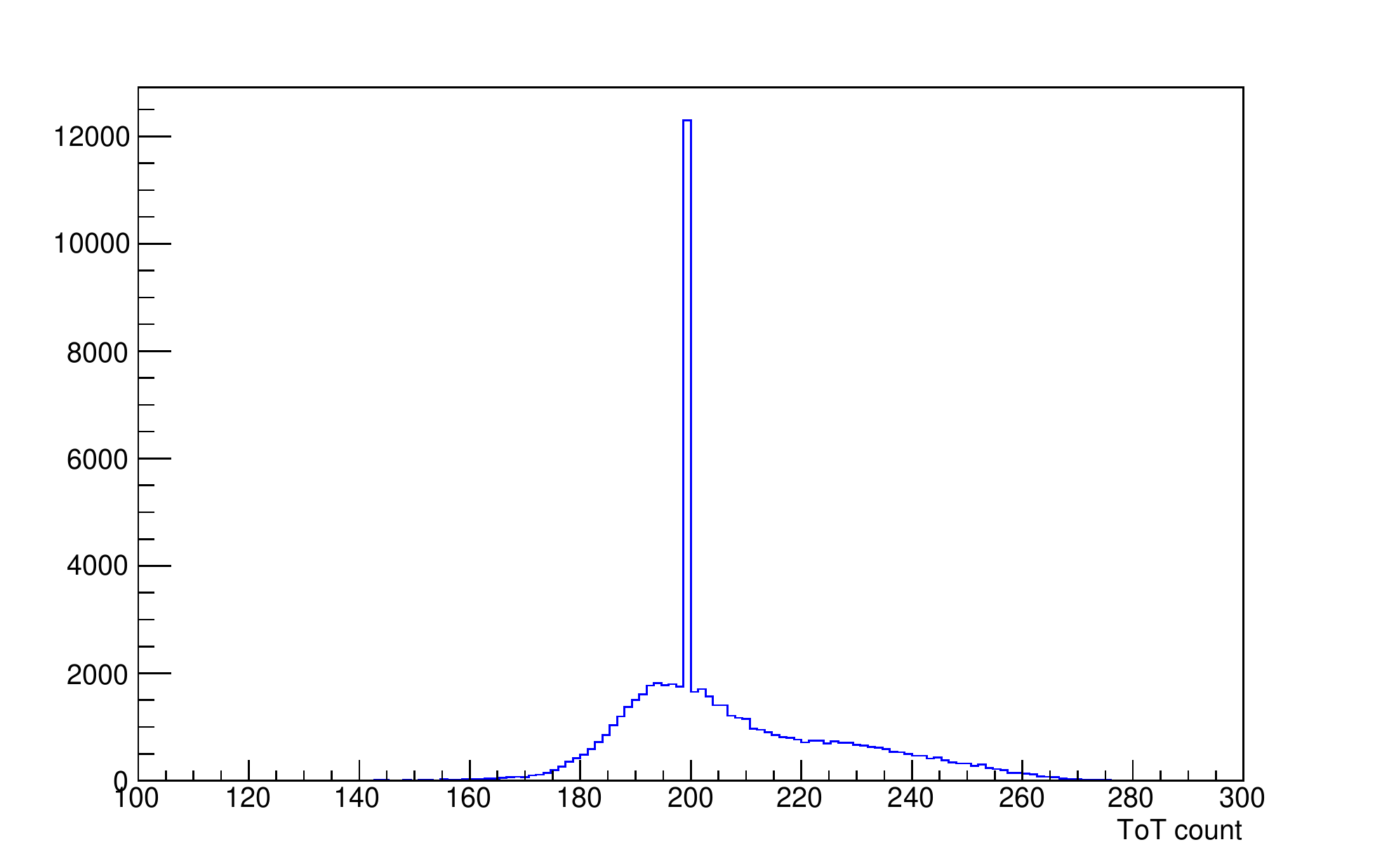} \\ b)}
\end{minipage}
\caption{\footnotesize Distribution of ToT values of the peak position in characteristic spectra of Rh (a) and Ta (b) for all pixels. Spikes in the distributions correspond to the peak positions in the summed spectra (these values are used for ``bad'' pixels).}
\label{PeakPos}
\end{figure}

The distributions of ToT values of the fitted peak positions in characteristic spectra of rhodium and tantalum for all pixels of the detector matrix are shown in Fig.~\ref{PeakPos}. These plots clearly illustrate shifts of the spectral peak from pixel to pixel that is the main motivation for per-pixel calibration. The ``spikes'' in the distributions are due to dead or improperly working pixels, for which $ ToT $ values of peak positions were set equal to the fitted mean values of the summed ToT spectra used in the generalized calibration.

\begin{figure}[!htp]
\center{\includegraphics[width=\linewidth]{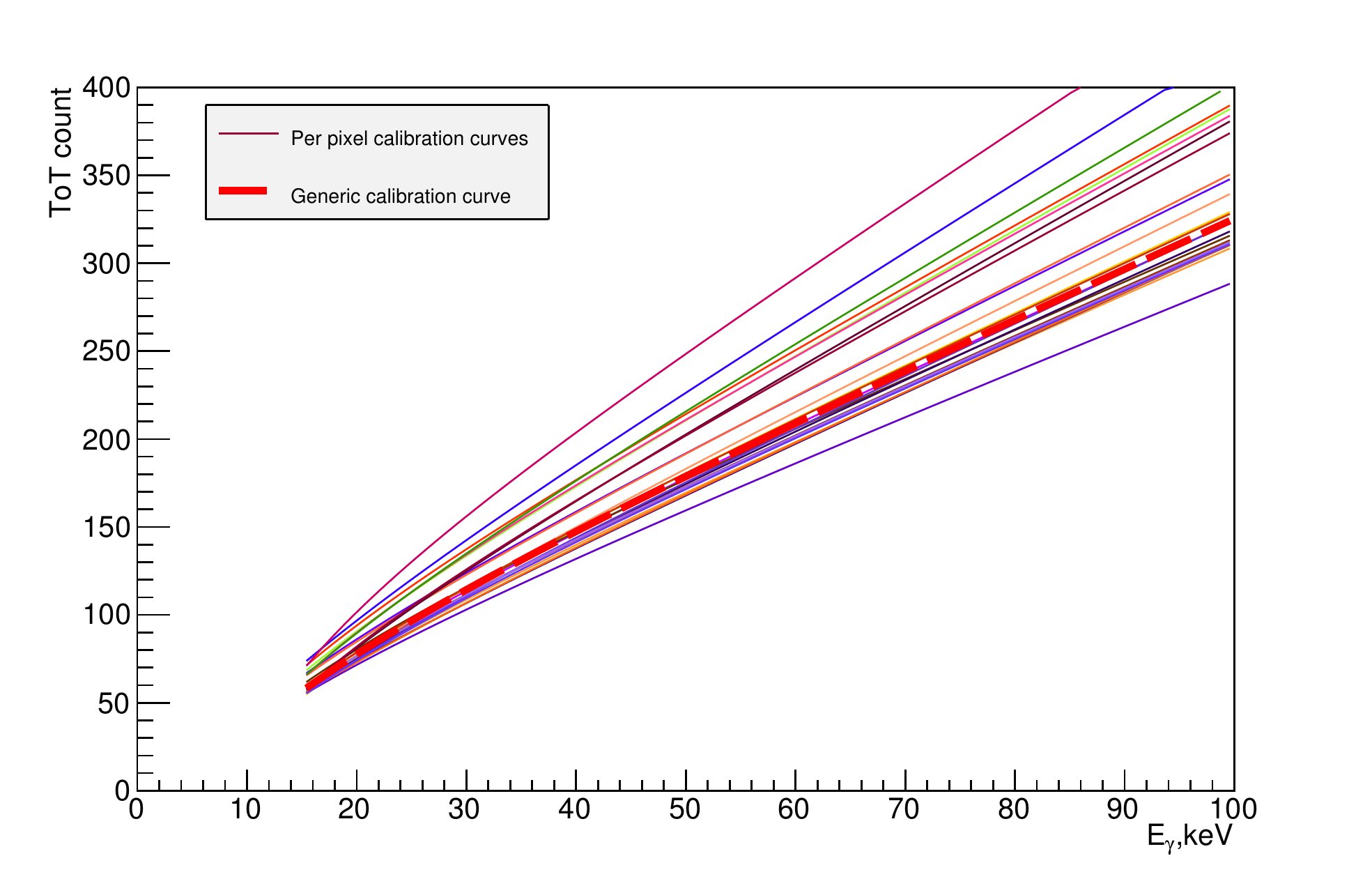}}
\caption{\footnotesize Per-pixel calibration curves for several pixels (thin lines) and generalized calibration curve (thick dashed line).}
\label{PxlClbCurve}
\end{figure}

Knowing how the $ ToT $ values change from peak to peak in pixels, it is possible to obtain calibration curves (\ref{eq.ToT(E)}) for each of them the same way as in the generalized calibration. In Fig.~\ref{PxlClbCurve} such per-pixel calibration curves for 32 pixels are presented in comparison with the generalized calibration curve. The number of ``bad'' fits decreases with increasing per pixel statistics, but there are always some pixels that either initially were dead as a result of assembly defects, or became corrupted during the detector operation. For most pixels \mbox{($\sim$ 85\%)} of the Timepix detector E11-W0110 the fitting of the spectra is performed without any problems. The result of the per-pixel calibration is four 256$\times$256 matrices with 65536 sets of parameters $ a $, $ b $, $ c $, $ t $ for the function~(\ref{eq.ToT(E)}).

\subsection{Comparison of per-pixel and generalized calibrations}

\begin{figure}[htb]
\begin{minipage}[h]{0.5\linewidth} 
\center{\includegraphics[width=1.0\linewidth]{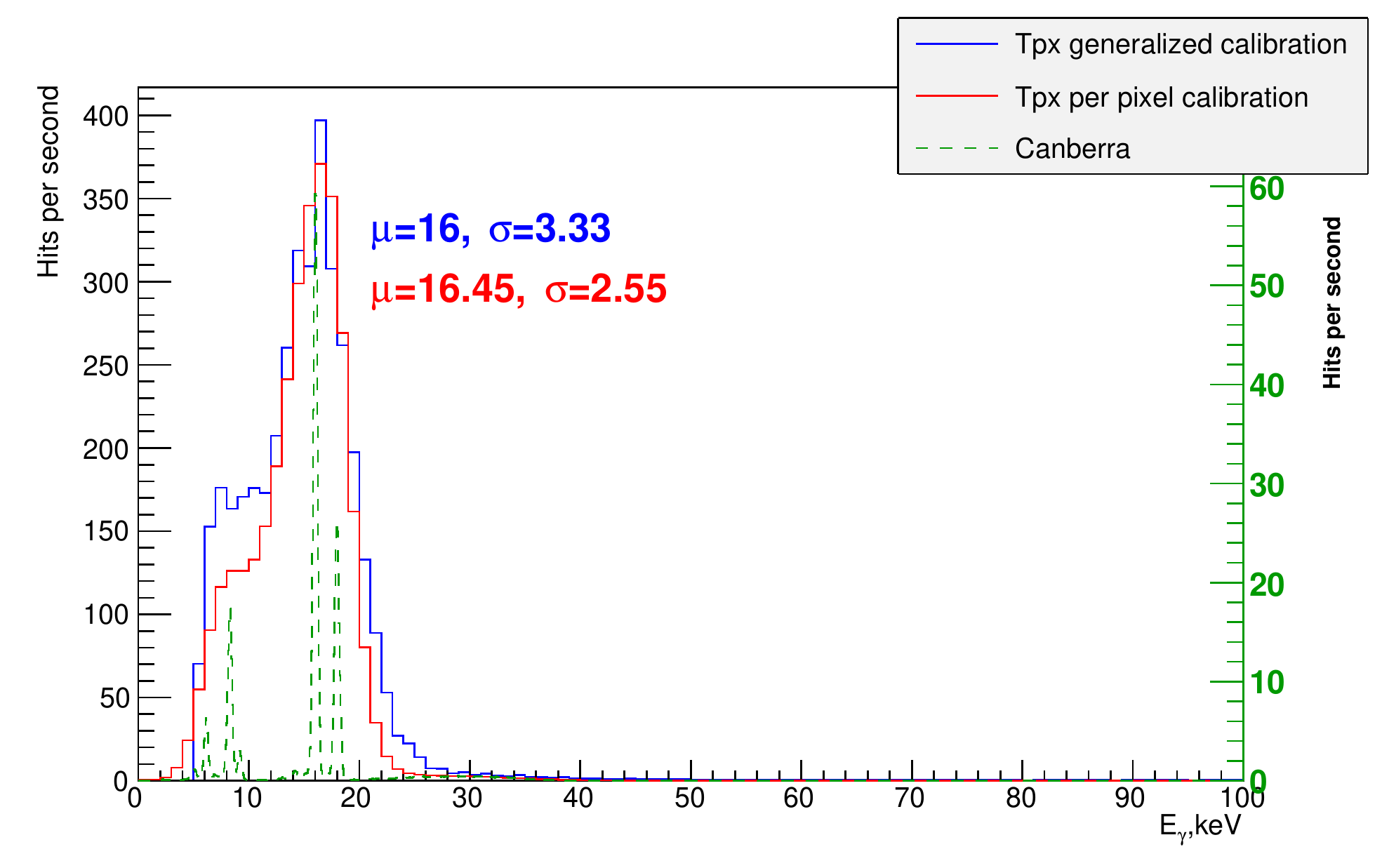} \\ a)} 
\end{minipage}
\hfill 
\begin{minipage}[h]{0.5\linewidth}
\center{\includegraphics[width=1.0\linewidth]{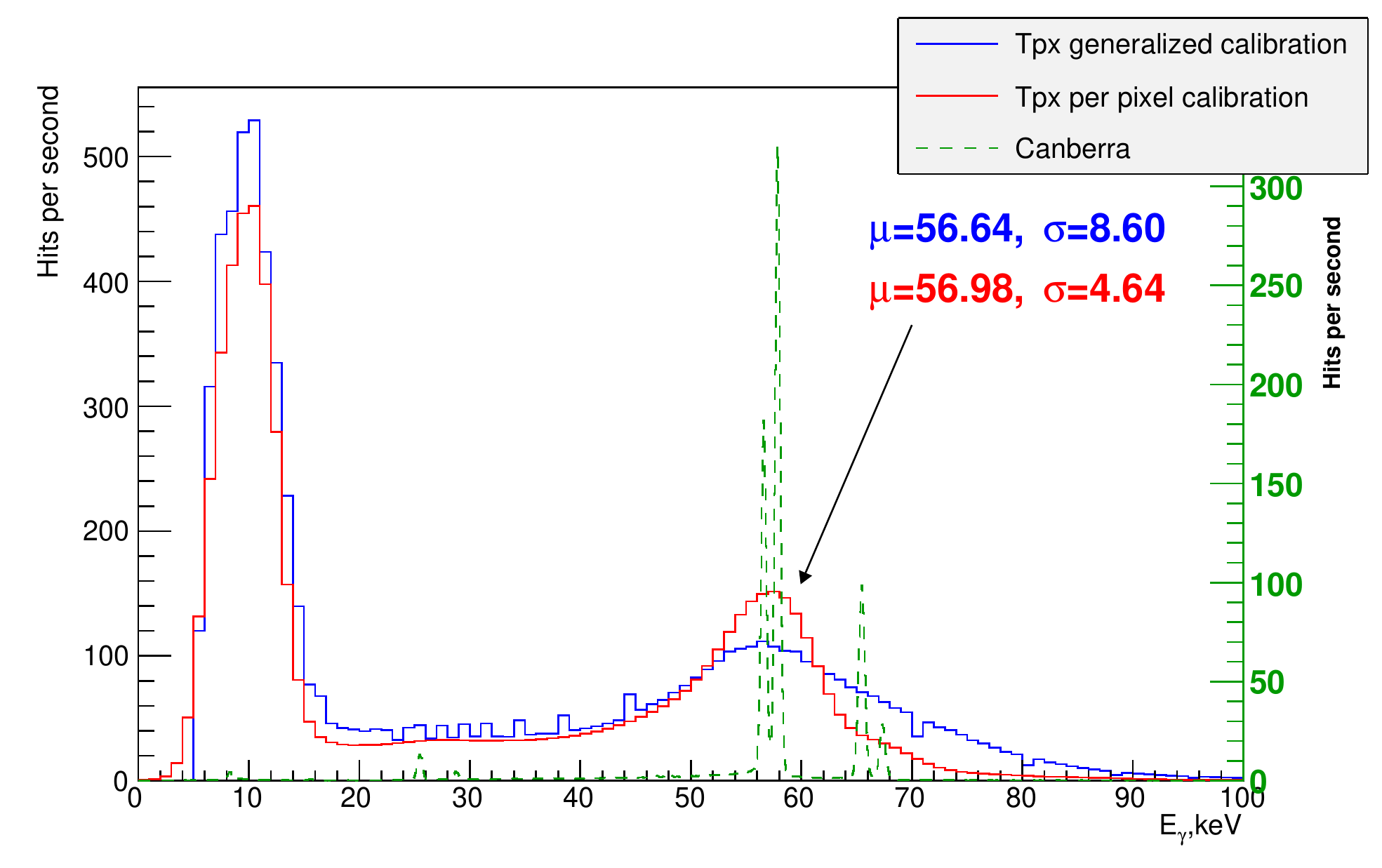} \\ b)}
\end{minipage}
\caption{\footnotesize Spectra of X-ray characteristic emission from Rh (a) and Ta (b) after per-pixel (red) and generalized (blue) calibrations.}
\label{GenPxlClb1}
\end{figure}

\begin{figure}[htb]
\begin{minipage}[h]{0.3\linewidth}
\center{\includegraphics[width=1.0\linewidth]{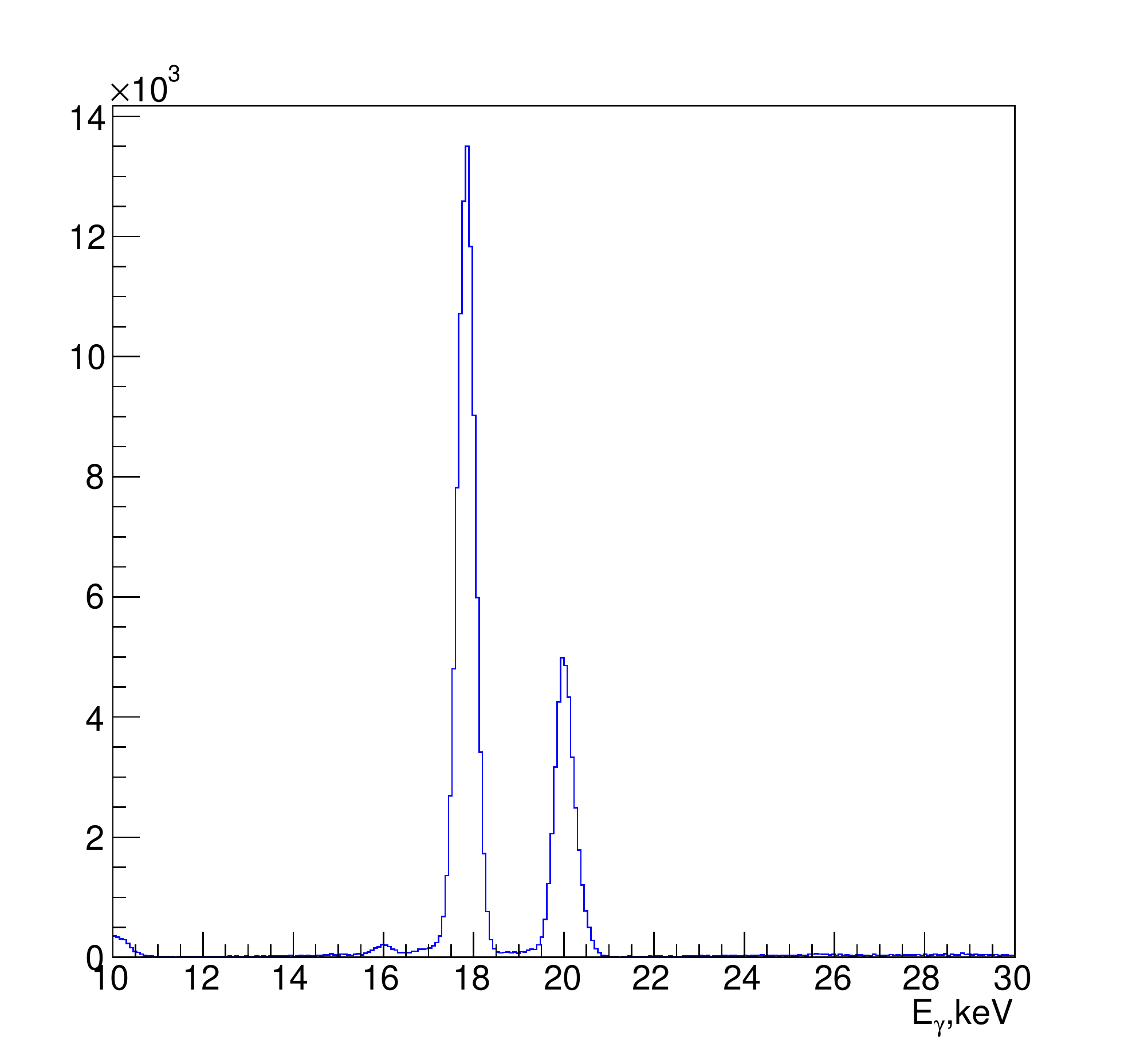} \\ a)}
\end{minipage}
\hfill
\begin{minipage}[h]{0.3\linewidth}
\center{\includegraphics[width=1.0\linewidth]{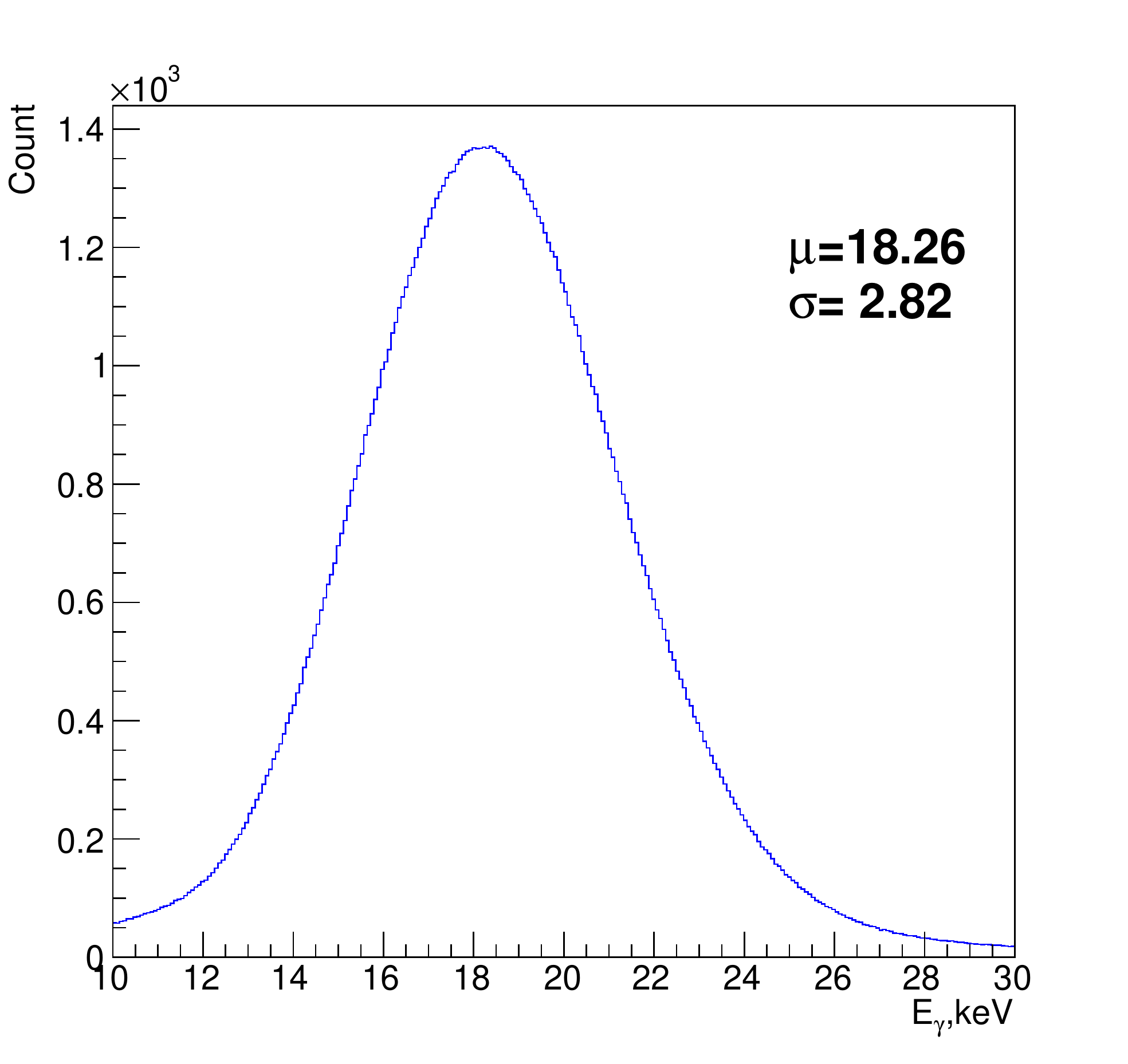} \\ b)}
\end{minipage}
\hfill
\begin{minipage}[h]{0.3\linewidth}
\center{\includegraphics[width=1.0\linewidth]{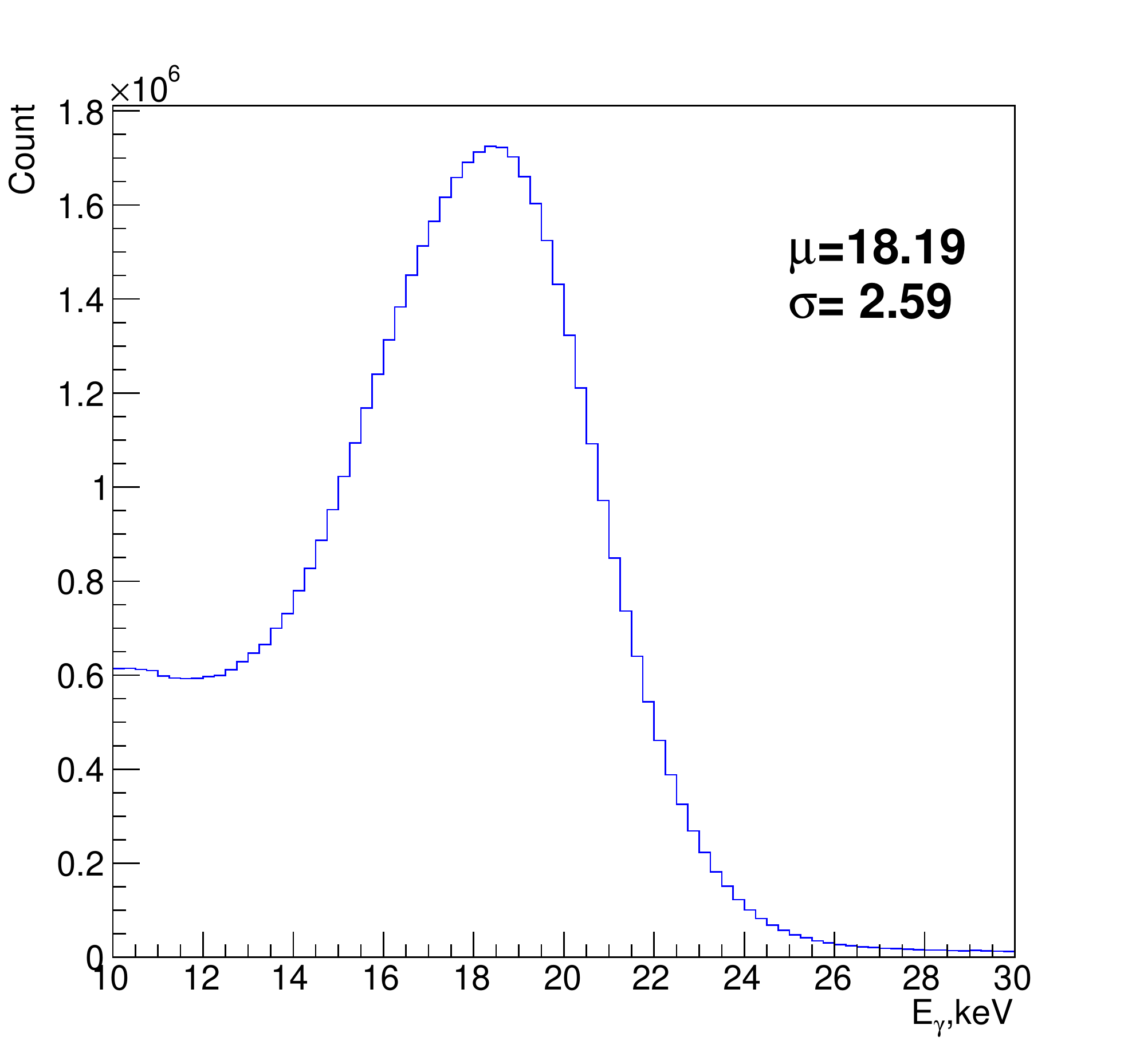} \\ c)}
\end{minipage}
\caption{\footnotesize Spectra of characteristic X-ray emission of Mo in the [10;30] keV range: a) measured with Canberra detector; b) modeled for GaAs Timepix detector; c) measured with GaAs Timepix detector after per-pixel calibration.}
\label{CompareSimReal}
\end{figure}

\begin{figure}[hbt]
\begin{minipage}[h]{0.5\linewidth}
\center{\includegraphics[width=1.0\linewidth]{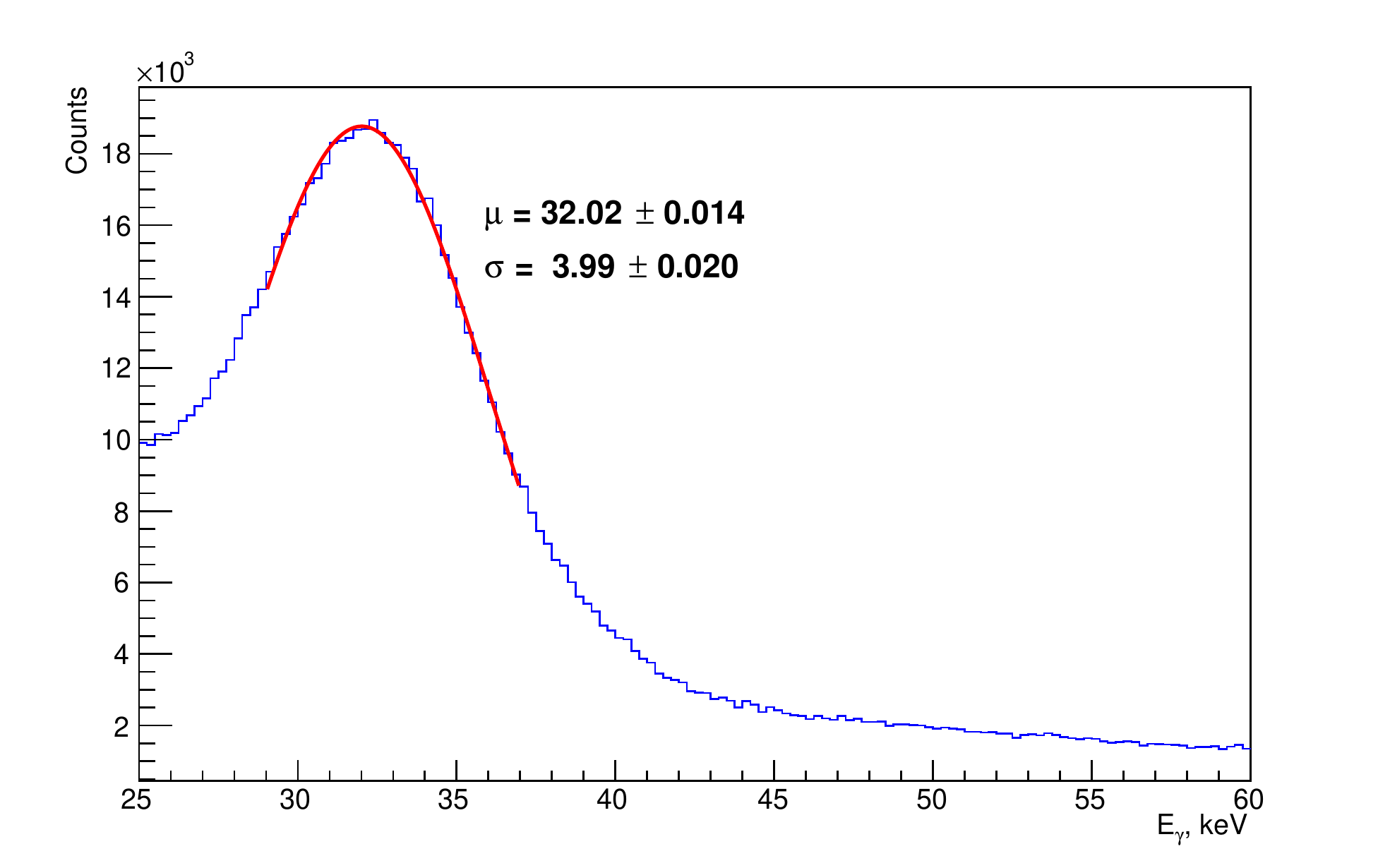} \\ a)}
\end{minipage}
\hfill
\begin{minipage}[h]{0.5\linewidth}
\center{\includegraphics[width=1.0\linewidth]{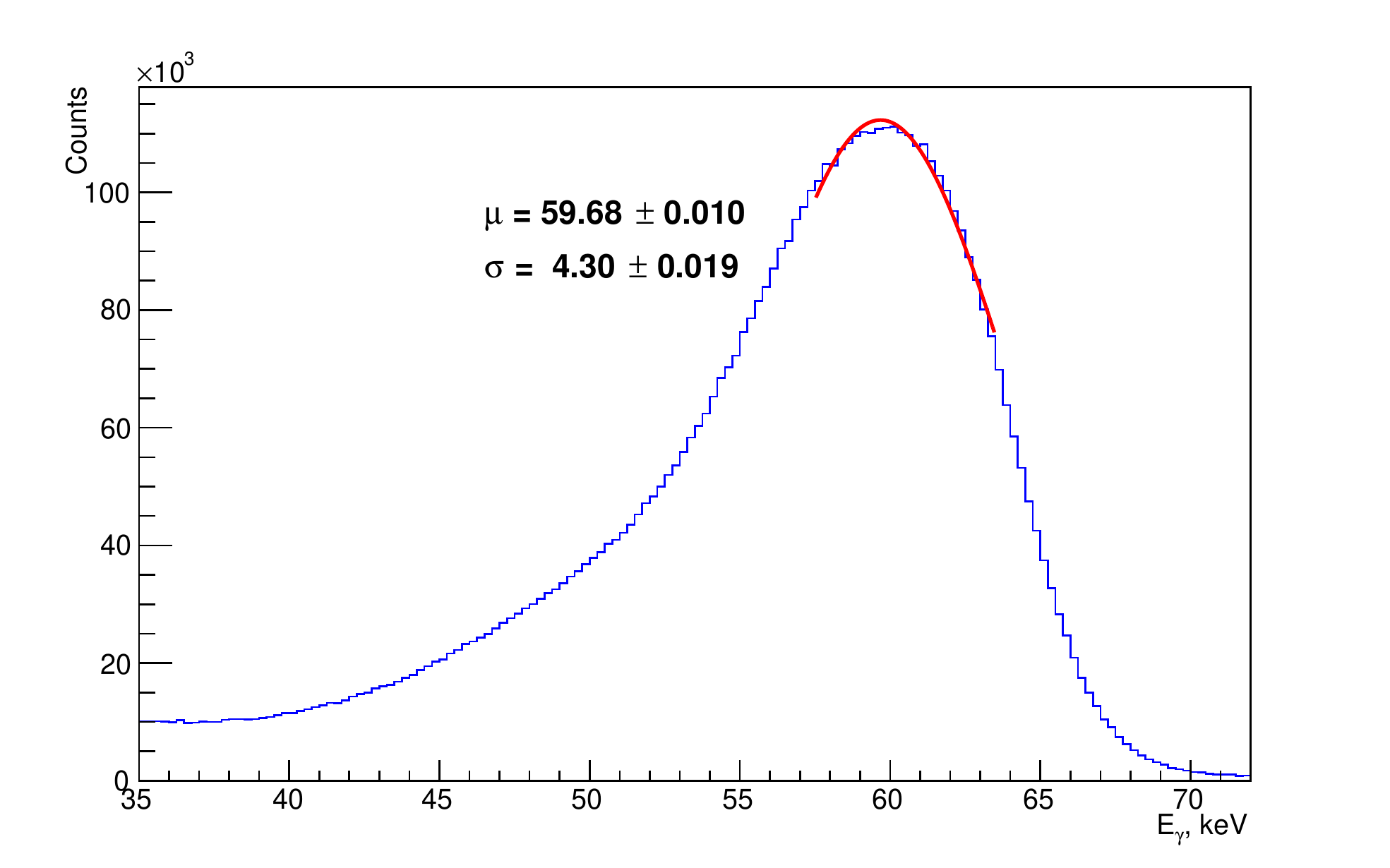} \\ b)}
\end{minipage}
\caption{\footnotesize Spectra of $\gamma$-radioactive sources measured by the Timepix detector after per-pixel calibration: a) $^{137}$Cs ($E_{\gamma}=32.18$ keV); b) $^{241}$Am ($E_{\gamma}=59.54$ keV).}
\label{RadSourcesSpectra}
\end{figure}

Following two types of calibration of the Timepix detector it is natural to study how the energy resolution $ \frac{\sigma}{E} $ behaves, depending on the type of calibration. For clarity, in Fig.~\ref{GenPxlClb1} the summed over all pixels\footnote{Everywhere in Figs.~\ref{GenPxlClb1}-\ref{RadSourcesSpectra} the spectra summed over all pixels are given.} spectra of characteristic emission of zirconium and tantalum, obtained after the generalized and per-pixel calibrations are shown.  From the figure it is clearly seen that per-pixel calibration significantly improves the resolution of the detector and the shape of the reconstructed spectra.

For quantitative comparison of the two calibrations, single peaks corresponding to the known energy in the calibrated spectra summed over all pixels were fitted with Gaussians. Then the ratios of the half-widths to the means for the two Gaussians were calculated. These values of the energy resolution of the detector $ \sigma_E = \sigma/E$, depending on the energy, along with the actual values of the energy calibration points are given in Table~\ref{CalibResol} (after two iterations of the Monte Carlo simulation, as described above).

\begin{table}[htbp]
\caption{\label{CalibResol}{Energy resolution of the GaAs:Cr Timepix detector E11-W0110 used in MC and measured with generalized and per-pixel calibrations.}}
\begin{center}
  \begin{tabular}{|c|c|c|c||c|c|c||c|c|c|c|c|c|} \hline
    & \multicolumn{3}{c||}{Monte Carlo} & \multicolumn{3}{|c||}{Generalized calibration} & \multicolumn{3}{c|}{Per-pixel calibration} \\ \cline{2-10}
    & $\sigma$ & $E$ & $\sigma / E$, \% & $\sigma$ & $E$ & $\sigma / E$, \% & $\sigma$ & $E$ & $\sigma / E$, \%  \\ \hline
    $^{40}$Zr & 2.782 & 16.50 & 16.9 & 3.332 & 15.99 & 20.8 & 2.545 & 16.45 & 15.5 \\
    $^{42}$Mo & 2.821 & 18.26 & 15.4 & 3.492 & 17.94 & 19.5 & 2.59 & 18.19 & 14.2 \\
    $^{45}$Rh & 2.792 & 20.93 & 13.3 & 3.661 & 20.55 & 17.8 & 2.562 & 20.87 & 12.3 \\
    $^{48}$Cd & 3.049 & 23.85 & 12.8 & 3.951 & 23.19 & 17.4 & 2.766 & 23.38 & 11.8 \\
    $^{49}$In & 3.037 & 24.92 & 12.2 & 4.245 & 24.56 & 17.3 & 2.825 & 24.63 & 11.4 \\
    $^{50}$Sn & 3.121 & 25.92 & 12.0 & 4.468 & 26.00 & 17.2 & 2.936 & 25.96 & 11.3 \\
    $^{73}$Ta & 5.655 & 56.91 & 9.9 & 8.602 & 56.64 & 15.2 & 4.641 & 56.98 & 8.1 \\ \hline
\end{tabular}
\end{center}
\end{table}

In Fig.~\ref{CompareSimReal}a the spectrum of the characteristic emission of molybdenum in the energy range [10,30] keV measured by the Canberra detector is shown. The modeled on its basis spectrum, accounting for the efficiency and resolution of the GaAs:Cr Timepix detector, is presented in Fig.~\ref{CompareSimReal}b. In Fig.~\ref{CompareSimReal}c the same spectrum measured with the Timepix detector after per-pixel calibration is shown. The presented spectra have peaks with similar values of the mean energy (within the error of the fit). It should be noted that the Monte Carlo simulation does not take into account the noise in the Timepix detector electronics and the charge sharing effect, leading to difference between the spectra at low energies.

As follows from Table~\ref{CalibResol}, performing the per-pixel calibration significantly improves the energy resolution of the Timepix detector. The reason for this is that the ToT spectra measured by each pixel, differ from each other. The difference between the maximum and minimum $ ToT $ values in a spectral peak, measured by different pixels, can be up to $ 6\sigma $. In the energy range 15-25 keV per-pixel calibration improves the energy resolution by a factor of 1.3-1.5 while at 57 keV energy the improvement is almost two-fold compared with the generalized calibration.

To assess the quality of the energy calibration of the Timepix detector, the $\gamma$-spectra of radioactive sources of cesium ($^{137}$Cs, $E_{\gamma}$=32.18 keV) and americium ($^{241}$Am, $E_{\gamma}$=59.54 keV) were measured. As can be seen from Fig.~\ref{RadSourcesSpectra}, the peak positions in the measured spectra differ from the table values by less than 0.5\%. with the energy resolution being 12.5\% for 32 keV and 7.2\% for 60 keV.

\section{Conclusions}

In this paper, a detailed description of the calibration procedure of Timepix pixel detectors is given. The procedure uses characteristic lines in various X-ray emission spectra that were precisely measured with the LEGe Canberra GL0515R spectrometer and convoluted with the resolution and efficiency of a GaAs:Cr sensor. The main advantage of using of characteristic X-ray emission spectra is the high rate of data collection (performing the same procedure with a set of reference radioactive $\gamma$-sources is possible but the required time increases by an order of magnitude). The accuracy of the obtained energy scale in the 15-100 keV range for single-pixel clusters is $\sim$0.5\%.

The use of per-pixel calibration allows to achieve a good energy resolution of the Timepix detector with GaAs:Cr sensor: $\sim$8\% at 60 keV and $\sim$13\% at 20 keV (for single-pixel clusters). Improvements in the quality of the calibration in the low energy part of the spectrum (5-15 keV) can be made by increasing the number of reference energy points used for the calibration curve. If variation of parameters of Timepix detectors is low, i.e. the detectors produce similar ToT spectra with the same DAC settings under stable temperature, the developed software allows to perform per-pixel calibration in mostly automated regime. This is especially important for systems with a large number of Timepix detectors.


                                                                                
\end{document}